\documentclass[osajnl,showpacs,10pt]{revtex4-2}
\usepackage{amsmath,amssymb,graphicx}
\usepackage{float}
\usepackage{multirow}
\DeclareMathOperator{\sech}{sech}
\usepackage[section]{placeins}

\begin{document}

\title{Families of Skyrmions in Two-Dimensional Spin-1/2 Systems}
\author{Amaria Javed, L.Al Sakkaf, and U. Al Khawaja\\
		Department of Physics, United Arab Emirates University,\\ P.O. Box
15551, Al-Ain, United Arab Emirates.\\ u.alkhawaja@uaeu.ac.ae}

\begin{abstract}
	We find Skyrmion-like topological excitations for a two-dimensional
	spin-1/2 system. Expressing the spinor wavefunction in terms of a
	rotation operator maps the spin-1/2 system to a Manakov system. We
	employ both analytical and numerical methods to solve the resulting
	Manakov system. Using a generalized similarity transformation, we reduce
	the two-dimensional Manakov system to the integrable one-dimensional
	Manakov system. Solutions obtained in this manner diverge at
	the origin. We employ a power series method to obtain an infinite
	family of localized and nondiverging solutions characterized by a
	finite number of nodes. A numerical method is then used to obtain a
	family of localized oscillatory solutions with an infinite number of
	nodes corresponding to a skyrmion composed of concentric rings with
	intensities alternating between the two components of the spinor. We investigate the stability of
	the skyrmion solutions found here by calculating their energy
	functional in terms of their effective size. It turns out that
	indeed the skyrmion is most stable when the phase difference between
	the concentric rings is $\pi$, i.e., alternating between spin up and spin down. Our results are also applicable to doubly polarized optical pulses.
\end{abstract}
\maketitle

\section{Introduction}\label{introduction section}

The vector non-linear Schr\"odinger equation (NLSE) describes spinor systems and the interaction between its field components. The model has also various applications in different areas of
physics, for instance, the propagation of electromagnetic waves with
arbitrary polarization in a self-focusing media \cite{1}, and the
evolution of waves in plasma \cite{2}. It is also shown \cite{3} that the
vector NLSE governs the average dynamics of
dispersion-managed solitons which are considered as a key element for optical communication. The soliton robustness to polarisation-mode dispersion has a strong dependence on both chromatic dispersion and soliton energy \cite{22}.
It is known that the two-component vector NLSE or the Manakov system is completely integrable and is solvable by the inverse scattering transform (IST) method. Recently, Manakov spatial solitons were observed in AlGaAs planar waveguides \cite{4}. More recently, the Si-based waveguides using similar phenomena are served as optical biosensors \cite{23}.
The similarity reductions of the 2D coupled NLSE have been studied by Lie's method. It is shown that the 2D coupled NLSE is reduced to the 1D-NLSE by the similarity transformations \cite{5}. The theoretical investigation for the evolution of and interaction between collective excitations in the two-dimensional NLSE was numerically studied by using shooting method and split-step Fourier method as well as the modulation instability method \cite{6}. The analytical bright one- and two-soliton solutions of the (2+1)-dimensional coupled NLSE under certain constraints were presented in Ref. \cite{7} by employing the Hirota method. Hirota method was also applied on the mixed-type solitons for a (2+1)-dimensional $N$-coupled nonlinear Schr\"odinger system in nonlinear optical-fiber communication \cite{8}.
The exact soliton solutions for the (2+1)-dimensional coupled higher-order NLSE in birefringent optical-fiber communication is given in \cite{9}.
The dynamical evolution of two-component Bose-Einstein condensates trapped in cylindrical well is numerically investigated by solving the coupled Gross-Pitaevskii equations and different numbers of unstable ring dark (gray) solitons were generated \cite{10}. The study of dark-bright (DB) ring solitons in two-component Bose-Einstein condensates is conducted in Ref. \cite{11}.
The Newton relaxation method was used in Ref. \cite{12} to obtain stationary discrete
vector solitons in two-dimensional nonlinear waveguide arrays. These results may also be applicable
to two-component Bose-Einstein condensates trapped in a two-dimensional optical lattice. The two dimensional
discrete solitons in optically induced nonlinear photonic lattices were observed in Ref. \cite{13}.
The homotopy analysis method was also used to solve cubic and coupled nonlinear Schr\"odinger
equations \cite{14}. The interaction of optical beams with arbitrary polarizations in self-focusing media is
studied in Ref. \cite{15} by using the direct scattering problem. Their physical schemes deal with spatial
solitons, and the dynamics is formally described by the initial value problem for the Manakov system. The ferromagnetic Bose-Einstein condensate allows for pointlike topological excitations, i.e., skyrmions \cite{16}. The stability of skyrmions in a fictitious spin-1/2 condensate is investigated in \cite{17}. The monopoles in an antiferromagnetic Bose-Einstein condensate and their static and dynamic properties were shown in Ref. \cite{18}. Topological protection of photonic mid-gap defect modes is demonstrated in Ref. \cite{19}. The description of topological phase transitions in photonic waveguide arrays is discussed in \cite{20}.\\
In particular, we are motivated to investigate the behaviour and stability of two-dimensional topological excitations in spin-1/2 system through a novel approach. We start with the calculation of rotation operator which is used to map the spin 1/2 system into a Manakov system that is considered as a model of wave propagation in fiber optics and provides the spin texture of skyrmions. The challenge is to solve the obtained 2D Manakov system, in order to find the nontrivial spin texture. We solved the 2D Manakov system through various analytical and numerical techniques. We used similarity transformation and found all solutions to diverge at the origin, $\rho = 0$. Then, we found nondiverging densities through power series method but with trivial textures. Finally, we used a numerical method to find nondiverging and nontrivial spin textures. The stability of these nondiverging and nontrivial  skyrmions is also investigated. We show that the two spin states (spin up and spin down) are in fact responsible for the stability of two-dimensional topological excitations.\\
This paper is organized as follows. In Sec.~\ref{2Dsky}  we calculate the spinor wavefunction and texture for all the possible cases of rotations. Mapping the spin-1/2 system to a
2D Manakov system is described in Sec.~\ref{map}. In Sec.~\ref{sol}, we solve the Manakov system to obtain nondiverging and nontrivial skrmions. We applied similarity transformation in Sec.~\ref{sim}, power series method in Sec.~\ref{ps} and numerical method in Sec.~\ref{num}. The stability of the nondiverging and nontrivial skyrmions is investigated in Sec.~\ref{stab}. Finally, we conclude by summarizing our main results in Sec.~\ref{con}.

\section{Two-dimensional skyrmions}\label{2Dsky}

A spinor wavefunction contains two degrees of freedom: total density ${n(\textbf{r},t)}$ and the spinor $\zeta(\bf{r})$ which has two components since we consider spin-1/2.
The total wavefunction is thus written as
\begin{eqnarray}
\Psi ({\bf{r}},t) = \sqrt{n({\bf{r}},t)}~\zeta(\bf{r}),
\end{eqnarray}
which obeys the NLSE
\begin{eqnarray}\label{NLS}
i\frac{\partial}{\partial t}\Psi({\bf{r}},t) = - \nabla^2
\Psi({\bf{r}},t) - \gamma\, |\Psi({\bf{r}},t)|^2 \Psi({\bf{r}},t).
\end{eqnarray}
The spin part of the wavefunction, $\zeta(\bf{r})$, can be parametrized by a rotation operator as
\begin{eqnarray}
\zeta({\bf{r}}) = \exp \Big\{ -\frac{i}{S}\bf{\Omega}(\bf{r}).{\bf{S}}\Big\}~ \zeta.
\end{eqnarray}
Here \textbf{S} is the spin matrix, ${\bf S}= \sigma_x \hat{x} +
\sigma_y\hat{y} + \sigma_z \hat{z}$, with $\sigma_x$, $\sigma_y$ and
$\sigma_z$ being the Pauli matrices. This operator amounts to a
rotation of the constant spin $\zeta$ around the vector
$\bf{\Omega}(\bf{r})$. Considering spherically symmetric spin
textures and restricting the general rotation operator to be around
the vector \textbf{r} by an angle of $\omega(\bf{r})$
gives $\bf{\Omega }(\bf{r}) = \omega (\bf{r})
\bf{\hat{r}}$ as depicted schematically in Fig.1a. Average spin at a
position $r$ is rotated by an angle $\omega(r)/S$ from its initial
orientation. An explicit form of $\omega(r)$ determines a specific
texture of the skyrmion. The constant spin $\zeta$ can be taken as
any of the eigenvectors of the Pauli spin matrices, namely

\begin{equation}
\zeta_x = \dfrac{1}{\sqrt{2}}
\begin{pmatrix}
1 \\
1
\end{pmatrix}; ~~~~\textrm{eigenstate for} ~ \sigma_x,
\end{equation}
\begin{equation}
\zeta_y = \dfrac{1}{\sqrt{2}}
\begin{pmatrix}
1 \\
i
\end{pmatrix}; ~~~~\textrm{eigenstate for}~ \sigma_y,
\end{equation}
\begin{equation}
\zeta_z =
\begin{pmatrix}
1 \\
0
\end{pmatrix}; ~~~~\textrm{eigenstate for}~ \sigma_z.
\end{equation}
The rotation operator can be reduced to a useful formula as:
\begin{equation}
\exp \Big\{ -{\frac{i}{S}\omega(r) \hat{\textbf{r}}.{\textbf{S}}}\Big\} =
{\rm \bf I} \cos[\omega(r)] -2 i( \hat{\bf{r}}.\textbf{S}) \sin[\omega(r)] ,
\end{equation}
where ${\rm \bf I}$ is the identity matrix. Using this formula, the
spinor wavefunction takes the form
\begin{eqnarray}\label{spinor function}
\Psi ({\bf{r}},t) = \sqrt{n({\bf{r}},t)}  \nonumber \\  \times \begin{pmatrix}
\cos[\omega(r)] - i \cos(\theta)\sin[\omega(r)] \\
\sin(\theta)\Big(-i \cos(\phi)+ \sin(\phi)\Big)\sin[\omega(r)]
\end{pmatrix},
\end{eqnarray}
where we have taken $\zeta = \begin{pmatrix}1\\0 \end{pmatrix}$. It is then straightforward to obtain the spin texture in terms of the average spin components
\begin{equation}<S_x> = \zeta^\dagger (r) S_x \zeta(r)\label{sxav},\end{equation}
\begin{equation}<S_y> = \zeta^\dagger (r) S_y \zeta(r)\label{syav},\end{equation}
\begin{equation}<S_z> = \zeta^\dagger (r) S_z \zeta(r)\label{szav}.\end{equation}

In the present work, we restrict the investigation to two-dimensional spin textures. To obtain a two-dimensional spin texture, we consider the three possible planes, namely $xz$-, $yz$-, and $xy$-planes. We consider the three possible initial spinors, namely $\zeta_x$, $\zeta_y$, and $\zeta_z$ and the three possible rotation axes, namely $x$-, $y$-, and $z$-axes. We consider also an interesting case with rotations in the $xy$-plane around the $\rho$-axis. Inspecting all possible cases, we found only three fundamentally and nontrivial different types of textures. The first is constructed by spins rotated around a fixed axis normal to the plane. The second is obtained when the spins are rotated around a fixed axis parallel to the plane. The third is obtained when spins are rotated around $\rho$ in the $xy$-plane. In the following we show the details for calculating the three spin textures.
\\
Considering rotations around $x$, $y$, $z$, or $\rho$-axis, we replace $\hat{r}$ by $\hat{x}$, $\hat{y}$, $\hat{z}$, or $\hat{\rho}$, respectively.\\
\begin{figure}
	\centering
		\includegraphics[scale=0.6]{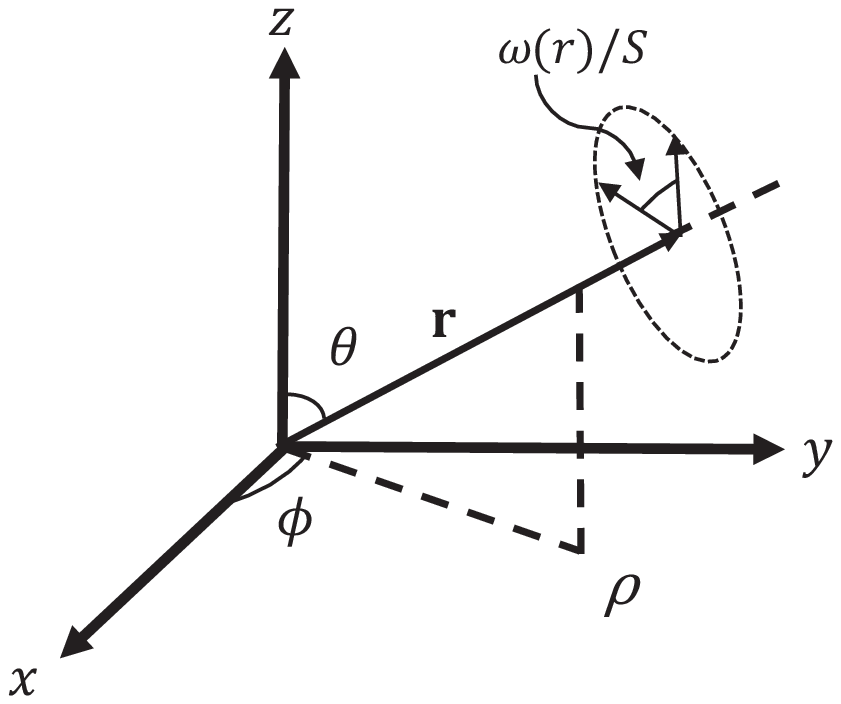}
		\includegraphics[scale=0.5]{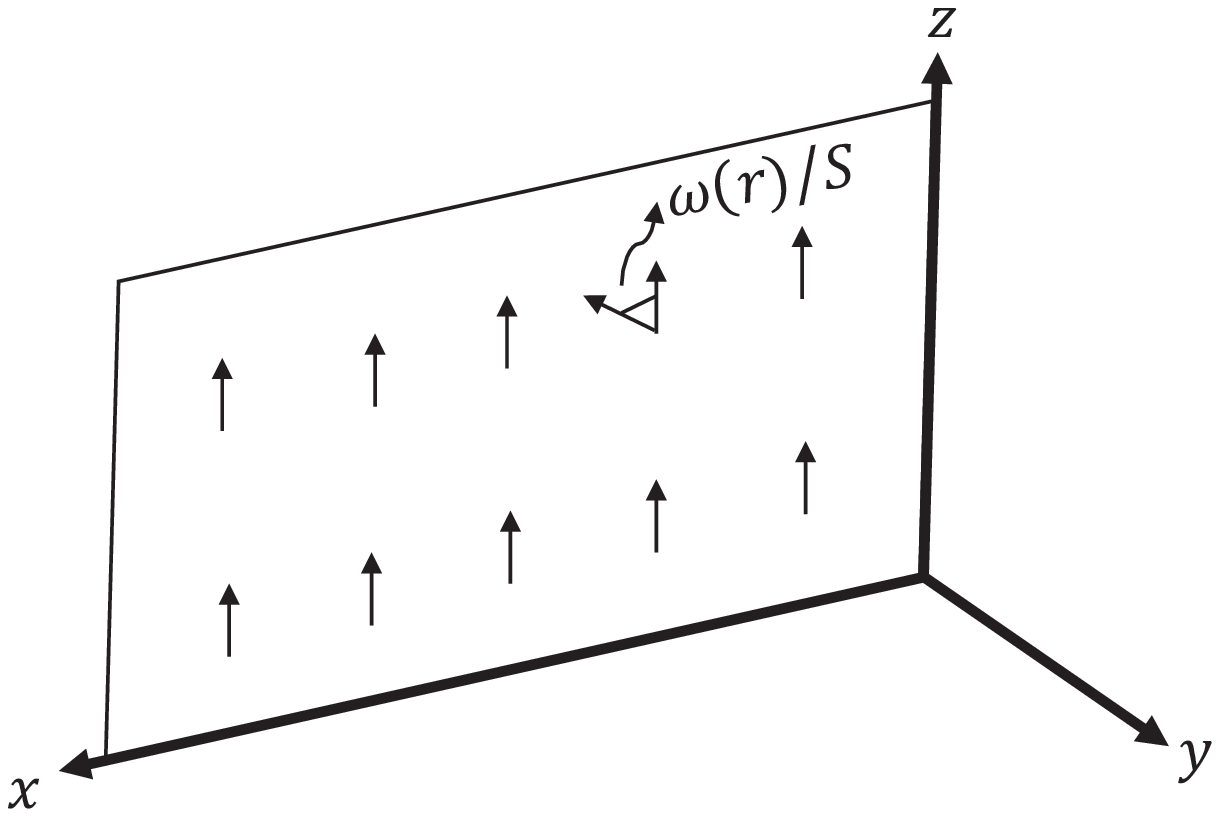}
	\caption{Schematic figure on left representing the action of the spin rotation operator for a maximally symmetric skyrmion while on right the rotation is around $y$-axis in $xz$-plane with initial spin along $z$-axis.}
	\label{fig:fig-xyz}
\end{figure}

\noindent\textbf{Rotations in the \textit{xz}-plane:}
We consider rotations in $xz$-plane with axis of rotation being the $x$-axis,
we choose  the initial orientation along \textit{z}-direction and hence use the
eigenvector of $S_{z}$, namely, $\zeta_z$,  for the operation.
The average spin components are given by
\begin{eqnarray}
\left(<S_x>, <S_y>, <S_z> \right)=\nonumber\\
\left( 0, - \dfrac{\sin[2\omega(\rho)]}{2},\dfrac{\cos[2\omega(\rho)]}{2}\right).
\end{eqnarray}
The spinor takes the form
\begin{equation}\zeta(\textbf{r})=
\begin{pmatrix}
\cos[\omega(\rho)] \\
- i\sin[\omega(\rho)]
\end{pmatrix}.
\end{equation}
This corresponds to spin rotations out of the plane, i.e., around an
axis parallel to the plane.
\\Considering rotations around the $y$-axis, the average spin components become
\begin{eqnarray}
\left(<S_x>, <S_y>, <S_z> \right)=\nonumber\\\left(  \dfrac{\sin[2\omega(\rho)]}{2}, 0,\dfrac{\cos[2\omega(\rho)]}{2}\right),
\end{eqnarray}
and the spinor becomes
\begin{equation}\zeta(\textbf{r})=
\begin{pmatrix}
\cos[\omega(\rho)] \\
\sin[\omega(\rho)]
\end{pmatrix}.
\end{equation}
This spin texture corresponds to spin rotations within the plane, i.e.,
around an axis perpendicular to the plane.
\\Considering the rotations around $z$-axis, we get
\begin{equation}\zeta(\textbf{r})=
\begin{pmatrix}
\cos[\omega(\rho)]  - i\sin[\omega(\rho)] \\
0
\end{pmatrix},
\end{equation}
and the average spin components are
\begin{equation}
\left( <S_x> , <S_y>, <S_z>\right)=(0,0,1/2),
\end{equation}
which is trivial case because it corresponds to spin rotations around the same axis along which the spins are aligned, and thus will be ignored.
\\The spinor and the average spin components for all the possible cases of rotations around fixed axes in $xz$-plane are listed in Table \ref{tab:xz-plane}. Considering other planes leads basically to only these two spin textures.
	\begin{table*}[!t]
	\begin{center}
		\begin{tabular}{ |c|c|c|c| }
			\hline
			\multicolumn{4}{|c|}{Rotations in $xz$-plane around fixed axes } \\
			\hline
			Axis of rotation & Initial Spin Orientation & $\zeta(\textbf{r})$ & $ \left(<S_x>,<S_y>,<S_z> \right)$ \\
			\hline
			\multirow{3}{6em}{$~~~~x$-axis $
				\begin{array}{l}
				~~~\theta = \pi/2,\\
				~~~\phi = 0
				\end{array}
				$} & $\zeta_x$ & $ \left(e^{-i\omega(\rho)}~\zeta_x\right) $ & $\left( \frac{1}{2}, 0, 0 \right)$ \\
			& $\zeta_y$ & $\begin{pmatrix}
			\dfrac{\cos[\omega(\rho)] + \sin[\omega(\rho)]}{\sqrt{2}} \\
			\dfrac{i(\cos[\omega(\rho)] -  \sin[\omega(\rho)])}{\sqrt{2}}
			\end{pmatrix}$ & $\left(0, \dfrac{\cos[2\omega(\rho)]}{2}, \dfrac{\sin[2\omega(\rho)]}{2}\right)$ \\
			& $\zeta_z$ & $\begin{pmatrix}
			\cos[\omega(\rho)] \\
			- i\sin[\omega(\rho)]
			\end{pmatrix}$ & $\left(0,  - \dfrac{\sin[2\omega(\rho)]}{2}, \dfrac{\cos[2\omega(\rho)]}{2}\right)$ \\
			\hline
			\multirow{3}{6em}{$~~~~y$-axis $
				\begin{array}{l}
				~~~ \theta = \pi/2,\\
				~~~\phi = \pi/2
				\end{array}
				$}  & $\zeta_x$ & $\begin{pmatrix}
			\dfrac{\cos[\omega(\rho)] -  \sin[\omega(\rho)]}{\sqrt{2}} \\
			\dfrac{\cos[\omega(\rho)] + \sin[\omega(\rho)]}{\sqrt{2}}
			\end{pmatrix}$ & $\left(\dfrac{\cos[2\omega(\rho)]}{2} , 0,  - \dfrac{\sin[2\omega(\rho)]}{2}\right)$ \\
			& $\zeta_y$ & $\left(e^{-i\omega(\rho)}~\zeta_y \right)$ & $\left(0, \frac{1}{2}, 0\right)$ \\
			& $\zeta_z$ & $\begin{pmatrix}
			\cos[\omega(\rho)] \\
			\sin[\omega(\rho)]
			\end{pmatrix}$ & $\left( \dfrac{\sin[2\omega(\rho)]}{2}, 0,  \dfrac{\cos[2\omega(\rho)]}{2}\right)$ \\
			\hline
			\multirow{3}{6em}{$~~~~z$-axis $
				\begin{array}{l}
				~~~\theta = 0,\\
				~~~\phi = \pi/2
				\end{array}
				$} & $\zeta_x$ & $\begin{pmatrix}
			\dfrac{\cos[\omega(\rho)] - i \sin[\omega(\rho)]}{\sqrt{2}} \\
			\dfrac{\cos[\omega(\rho)] + i \sin[\omega(\rho)]}{\sqrt{2}}
			\end{pmatrix}$ & $\left(\dfrac{\cos[2\omega(\rho)]}{2}, \dfrac{\sin[2\omega(\rho)]}{2}, 0 \right)$ \\
			& $\zeta_y$ & $\begin{pmatrix}
			\dfrac{\cos[\omega(\rho)] - i \sin[\omega(\rho)]}{\sqrt{2}} \\
			\dfrac{i\cos[\omega(\rho)] -  \sin[\omega(\rho)]}{\sqrt{2}}
			\end{pmatrix}$ & $\left( - \dfrac{\sin[2\omega(\rho)]}{2}, \dfrac{\cos[2\omega(\rho)]}{2}, 0 \right)$ \\
			& $\zeta_z$ & $\left(e^{-i\omega(\rho)}~\zeta_z \right)$ & $\left(0, 0, \frac{1}{2}\right)$ \\
			
			\hline
		\end{tabular}
		\caption {
			All possible cases of rotations in the $xz$-plane around three possible fixed axes of rotation, $x$-, $y$-, and $z$-axis, with three possible initial spin directions, $\zeta_x$, $\zeta_y$, $\zeta_z$. The spinor wavefunction,  $\zeta({\bf r})$, and the average spin components, $<S_x>$, $<S_y>$ and $<S_z>$, are calculated using Eqs. (\ref{spinor function},
			\ref{sxav},\ref{syav},\ref{szav}).
		} \label{tab:xz-plane}
	\end{center}
	\end{table*}

\noindent	\textbf{Rotations in the \textit{xy}-plane around $\bf{\rho}$ :}
	We consider the rotations around $\rho = \sqrt{x^2 + y^2}$. Since the axis of rotation changes with $\phi$, the spinor components and the texture demand also on $\phi$, as listed in Table  \ref{tab:xy-plane}.
	\begin{table*}[!t]
	\begin{center}
		\vspace*{1.3cm}
		\begin{tabular}{ |c|c|c|c| }
			\hline
			\multicolumn{4}{|c|}{Rotations in $xy$-plane around $\rho$ } \\
			\hline
			Axis of rotation & Initial spin & $\zeta(\textbf{r})$ & $\left(<S_x>,<S_y>,<S_z>\right)$ \\
			\hline
			\multirow{3}{6em}{$\rho=~\sqrt{x^2+y^2}$\\ $~~(\theta = \pi/2)$} & $\zeta_x$ & $\begin{pmatrix}
			\dfrac{\cos[\omega(\rho)] - i e^{-i\phi} \sin[\omega(\rho)]}{\sqrt{2}} \\
			\dfrac{\cos[\omega(\rho)] - i  e^{i\phi} \sin[\omega(\rho)]}{\sqrt{2}}
			\end{pmatrix}$ & $\left( \dfrac{\cos^2[\omega(\rho)]+ \cos(2\phi)\sin^2[\omega(\rho)]}{2},  \dfrac{\sin(2\phi)\sin^2[\omega(\rho)]}{2}, - \dfrac{\sin(\phi)\sin[2\omega(\rho)]}{2}\right)$  \\
			& $\zeta_y$ & $\begin{pmatrix}
			\dfrac{\cos[\omega(\rho)] + e^{-i\phi} \sin[\omega(\rho)]}{\sqrt{2}} \\
			\dfrac{i(\cos[\omega(\rho)] - e^{i\phi} \sin[\omega(\rho)])}{\sqrt{2}}
			\end{pmatrix}$ & $ \left(\dfrac{\sin(2\phi)\sin^2[\omega(\rho)]}{2},  \dfrac{\cos^2[\omega(\rho)]- \cos(2\phi)\sin^2[\omega(\rho)]}{2},  \dfrac{\cos(\phi)\sin[2\omega(\rho)]}{2}\right)$ \\
			& $\zeta_z$ &$\begin{pmatrix}
			\cos[\omega(\rho)]  \\
			e^{-i (\frac{\pi}{2} - \phi)} \sin[\omega(\rho)]
			\end{pmatrix}$ & $\left(\dfrac{\sin(\phi)\sin[2\omega(\rho)]}{2}, - \dfrac{\cos(\phi)\sin[2\omega(\rho)]}{2}, \dfrac{\cos[2\omega(\rho)]}{2}\right)$  \\
			\hline
		\end{tabular}
		\caption {All possible cases of rotations in the $xy$-plane around $\rho$ with three possible initial spin directions, $\zeta_x$, $\zeta_y$, $\zeta_z$. The spinor wavefunction,  $\zeta({\bf r})$, and the average spin components, $<S_x>$, $<S_y>$ and $<S_z>$, are calculated using Eqs. (\ref{spinor function},
			\ref{sxav},\ref{syav},\ref{szav}).} \label{tab:xy-plane}
	\end{center}
	\end{table*}

\section{Mapping the spin-1/2 system to a 2D Manakov system}\label{map}
We have shown in the previous section that the spinor wavefunction
of a skyrmion can be written in a specific form that corresponds to
spin rotations. There are many such specific forms depending on the
plane at which the spins are located in, the axis of spin rotations,
and the initial spin orientation, as summarized by Tables 1 and 2.
This procedure is effectively a change of variables amounting to a
change in the representation from the spinor components, $\psi_1$
and $\psi_2$, to the total density $n$ and angle of rotation
$\omega/S$.
\\Considering one of these specific cases, namely spinors restricted
to the $xz$-plane with spin rotations around the $y$-axis as shown
schematically in Fig. \ref{fig:fig-xyz}b, the spinor becomes
\begin{eqnarray}\label{spinor}
\begin{pmatrix}
\psi_1(\rho,\phi,t)\\
\psi_2(\rho,\phi,t)
\end{pmatrix}
= \sqrt{n(\rho,t)}
\begin{pmatrix}
e^{il_1\phi} \cos[\omega(\rho)]\\
e^{il_2\phi} \sin[\omega(\rho)]
\end{pmatrix},
\end{eqnarray}
where $\rho=\sqrt{x^2+z^2}$ and $\phi$ is the angle between $\rho$ and the $z$-axis.
We have added the phase operator $e^{il\phi}$ to allow for non-zero angular momentum of any of the two components.
This accounts for an acquired phase while spins are rotated.
This spin-1/2 system is then mapped to a 2D Manakov system obtained
by substituting the spinor \eqref{spinor} in the NLSE, \eqref{NLS},\\
\begin{eqnarray}\label{2DMS}
i\begin{pmatrix}
\psi_1(\rho,\phi,t)\\
\psi_2(\rho,\phi,t)
\end{pmatrix}_t &=&  -\begin{pmatrix}
\psi_1(\rho,\phi,t)\\
\psi_2(\rho,\phi,t)
\end{pmatrix}_{\rho\rho} \nonumber\\ &-& \frac{1}{\rho}  \begin{pmatrix}
\psi_1(\rho,\phi,t)\\
\psi_2(\rho,\phi,t)
\end{pmatrix}_\rho \nonumber\\ &-& \frac{1}{\rho^2}\begin{pmatrix}
\psi_1(\rho,\phi,t)\\
\psi_2(\rho,\phi,t)
\end{pmatrix}_{\phi\phi} \nonumber\\ &-& \gamma ( |\psi_1(\rho,\phi,t)|^2 + |\psi_2(\rho,\phi,t)|^2 )\nonumber\\&\times& \begin{pmatrix}
\psi_1(\rho,\phi,t)\\
\psi_2(\rho,\phi,t)
\end{pmatrix}.
\end{eqnarray}
The problem then reduces to solving this system, which we describe in the next
section. The solutions $\psi_1(\rho,\phi,t)$ and $\psi_2(\rho,\phi,t)$
can then be used in \eqref{spinor} to obtain two coupled equations
for $n(\rho,t)$ and $\omega(\rho)$. Solving these equations gives
the texture of the skyrmion through $<S_x>$, $<S_y>$, and $<S_z>$, as well as its density
profile, $n(\rho,t)$.

\section{Solving the Manakov system}\label{sol}

We present different methods of solving the Manakov
system \eqref{2DMS} in order to generate the non-trivial spin
texture. All methods mentioned below are well known and powerful techniques in analytical and numerical analysis but our desired results are achieved by the numerical technique described in Sec.~\ref{num}. The other methods paved the way for developing the numerical technique on a trial function, so we include them in this section. At first, we attempt to map the 2D Manakov system
\eqref{2DMS} into the 1D Manakov system which is integrable with
many known solutions. While this leads to nontrivial skyrmion
textures, the corresponding spinor densities diverge at $\rho = 0$.
As an alternative approach we employ a power series method to find
well-behaved spinor densities. However, the associated skyrmion
texture turns out to be trivial for such a case. Finally,
well-behaved spinor densities with nontrivial skyrmion textures are
obtained by employing a trial function that takes into account the
spin texture of a specific case of rotation as detailed in Tables
\ref{tab:xz-plane}  and \ref{tab:xy-plane},
and then solving numerically the NLSE for $n(\rho,t)$ and $\omega(\rho)$.

\subsection{Similarity transformation}\label{sim}

At first, we transform the 2D Manakov system into the fundamental 1D
Manakov system via a simple similarity transformation. This will
enable us then to find the new solutions of 2D Manakov system by
using all known solutions of the fundamental 1D Manakov system. We
start with the simplest case for the solution of 2D Manakov system,
namely, the cylindrically symmetric solution. As there is no $\phi$
dependence in this case, corresponding to $l_1=l_2=0$, we are left
with
\begin{eqnarray}\label{MS}
i\begin{pmatrix}
\psi_1(\rho,t)\\
\psi_2(\rho,t)
\end{pmatrix}_t = -\begin{pmatrix}
\psi_1(\rho,t)\\
\psi_2(\rho,t)
\end{pmatrix}_{\rho\rho} -\frac{1}{\rho}  \begin{pmatrix}
\psi_1(\rho,t)\\
\psi_2(\rho,t)
\end{pmatrix}_\rho \nonumber\\ - \gamma ( |\psi_1(\rho,t)|^2 + |\psi_2(\rho,t)|^2 ) \begin{pmatrix}
\psi_1(\rho,t)\\
\psi_2(\rho,t)
\end{pmatrix}.
\end{eqnarray}
To reduce this system into the integrable 1D Manakov system we apply the following simple transformation
\begin{equation}\label{simple}
\psi_{1,2} = \rho^n ~ \Phi_{1,2}
\end{equation}
to the above 2D Manakov system \eqref{MS}. The system then reduces to the fundamental 1D Manakov system for $n= - 1/2$
\begin{equation}
\psi_{1,2} = \dfrac{1}{\sqrt{\rho}} ~ \Phi_{1,2}.
\end{equation}\\

For all solutions of the 1D Manakov system, $\Phi_{1,2}$, that are finite at $\rho = 0$,
the solutions of the 2D Manakov system $\psi_{1,2}$  diverge at  $\rho = 0$.
This applies to all known solutions of the 1D Manakov system which we have used in
Appendix \ref{appa}, except the solution $(\Phi_1, \Phi_2) \sim (\tanh(\rho), \sech(\rho))$. The $\tanh(\rho)$ part of this particular solution is zero at $\rho = 0$, and thus $\psi_1 \sim \Phi_1 / \sqrt{\rho}$ does not diverge at $\rho = 0$. However, the other component $\psi_2 \sim \sech(\rho) / \sqrt{\rho}$ diverges at $\rho = 0$. For all other solutions, both components diverge at $\rho = 0$.\\
\\To establish the link between the solutions of the 1D and 2D Manakov systems in a rigorous manner, we consider the following most general form of a similarity transformation
\begin{flalign}\label{source}
p_1~ \Bigg[ i ~ \psi_{1_t} + b_{11} \Big[ \psi_{1_{\rho\rho}} + \frac{1}{\rho}\psi_{1_\rho} \Big] + \Big[ b_{12}|\psi_1|^2 + \nonumber\\ b_{13}|\psi_2|^2 \Big]\psi_1
+ \Big[ b_{14r} + i b_{14i}\Big]\psi_1 \Bigg] = 0, \nonumber\\
p_2~ \Bigg[ i ~ \psi_{2_t} + b_{21} \Big[ \psi_{2_{\rho\rho}}+ \frac{1}{\rho}\psi_{2_\rho} \Big] + \Big[ b_{22}|\psi_1|^2 + \nonumber\\ b_{23}|\psi_2|^2 \Big]\psi_2
+ \Big[ b_{24r} + i b_{24i}\Big]\psi_2 \Bigg] = 0,
\end{flalign}
where $p_1$, $p_2$, $b_{11}$, $b_{21}$, $b_{12}$, $b_{22}$, $b_{13}$, $b_{23}$, $ b_{14r}$, $b_{24r}$, $b_{14i}$, and $b_{24i}$ are all functions of $(\rho,t)$, and are arbitrary real coefficients. We apply the following transformation on the system \eqref{source}
\begin{eqnarray}\label{trans}
\Psi_1 (\vec{r},t) = A(\rho,t) ~e^{i B_1(\rho,t)}~F[ P (\rho,t),T(\rho,t)] , \nonumber \\
\Psi_2 (\vec{r},t) = A(\rho,t) ~e^{i B_2(\rho,t)}~G[P(\rho,t),T(\rho,t)].
\end{eqnarray}
Here, $A(\rho,t)$, $B_1(\rho,t)$, $B_2(\rho,t)$, $P(\rho,t)$, and
$T(\rho,t)$ are all defined as real functions. Substituting
\eqref{trans} in \eqref{source} and requiring the resulting
equations to take the form of the following fundamental Manakov
system
\begin{eqnarray}\label{FMS}
i F_t(P,T) + a_{11} F_{\rho\rho}(P,T) + \Big[ a_{12}|F(P,T)|^2 + \nonumber\\ a_{13}|G(P,T)|^2 \Big]F(P,T) =0, \nonumber \\
i G_t(P,T) + a_{21} G_{\rho\rho}(P,T) + \Big[ a_{22}|F(P,T)|^2 + \nonumber\\
a_{23}|G(P,T)|^2 \Big]G(P,T) =0,
\end{eqnarray}
gives a set of equations for the unknown functions. Solutions of
these equations are relegated to Appendix \ref{appb}. We listed few
solutions for the 2D Manakov system obtained using this approach in
Appendix \ref{appa}. Here again, we end up with the solutions having
divergences at $\rho = 0$ and therefore they will be discarded for
no physical significance. Seeking solutions which are well-behaved
at $\rho=0$, we employ in the next section an Iterative Power Series
(IPS) method \cite{21}.

\subsection{Power series method}\label{ps}

We  apply the IPS method with a stationary solution given by
\begin{eqnarray}
\psi_1(\rho,t)=Z_1(\rho)\,e^{i\,\alpha_1\,t},\nonumber \\
\psi_2(\rho,t)=Z_2(\rho)\,e^{i\,\alpha_2\,t},
\end{eqnarray}
where $Z_1(\rho)$ and $Z_2(\rho)$ are real functions and $\alpha_1$ and $\alpha_2$ are arbitrary real constants. Using this solution, Eq. (\ref{MS}) renders to the following ordinary differential equations
\begin{eqnarray}\label{ode}
\frac{1}{\rho}\,Z_1^\prime(\rho)+Z_1^{\prime\prime}(\rho)+Z_1(\rho)\left[\gamma\,Z_2^2(\rho)-\alpha_1\right] \nonumber\\+\gamma\,Z_1^3(\rho)=0,\nonumber\\
\frac{1}{\rho}\,Z_2^\prime(\rho)+Z_2^{\prime\prime}(\rho)+Z_2(\rho)\left[\gamma\,Z_1^2(\rho)-\alpha_2\right]\nonumber\\+\gamma\,Z_2^3(\rho)=0.
\end{eqnarray}
In the following, we give a brief algorithm description of the IPS method for obtaining a convergent power series solution to (\ref{ode}):
\begin{enumerate}
	\item Expand  $Z_1(\rho)$ and $Z_2(\rho)$ in power series around an arbitrary real initial point $\rho_0$:\\ \\
	$Z_1(\rho)=a_0+a_1\,(\rho-\rho_0)+\sum_{n=2}^{n_{max}} a_n(\rho-\rho_0)^n$,\\ \\
	$Z_2(\rho)=b_0+b_1\,(\rho-\rho_0)+\sum_{n=2}^{n_{max}} b_n(\rho-\rho_0)^n$.
	\item Set initial values $\{a_0, a_1\}$ and
	$\{b_0, b_1\}$ for $Z_1(\rho)$ and $Z_2(\rho)$, respectively.
	\item Substitute in (\ref{ode}) to obtain the recursion relation for $a_n$ and $b_n$ in terms of $a_0$, $a_1$, $b_0$, and $b_1$.
	\item Calculate $Z_1(\Delta)$, ${Z_1}^\prime(\Delta)$,  $Z_2(\Delta)$,  and ${Z_2}^\prime(\Delta)$, where $\Delta=(\rho-\rho_0)/I$  and $I$ is an integer larger than 1.
	\item Assign: $a_0=Z_1(\Delta)$, $a_1={Z_1}^\prime(\Delta)$,  $b_0=Z_2(\Delta)$, and $b_1={Z_2}^\prime(\Delta)$.
	\item Obtain $a_n$ and $b_n$ in terms of $a_0$, $a_1$, $b_0$, and $b_1$.
	\item Repeat steps 2-6 $I$ times.
	\item At the $I$th step, $a_0$ will correspond to the power series of $Z_1(\rho)$ and   $b_0$ will correspond to the power series of $Z_2(\rho)$.
\end{enumerate}
\begin{figure}[h!]
	\centering
		\includegraphics[scale=.7]{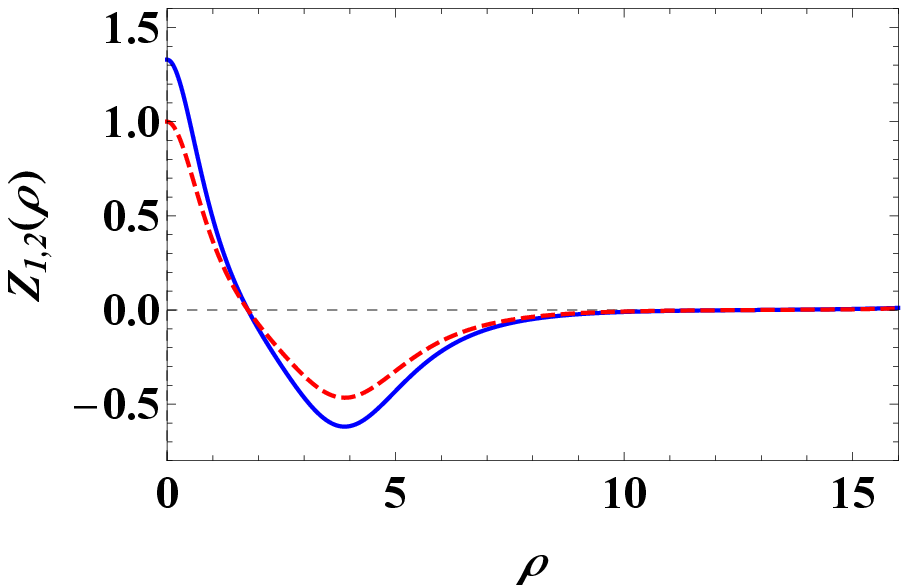}
		\includegraphics[scale=.7]{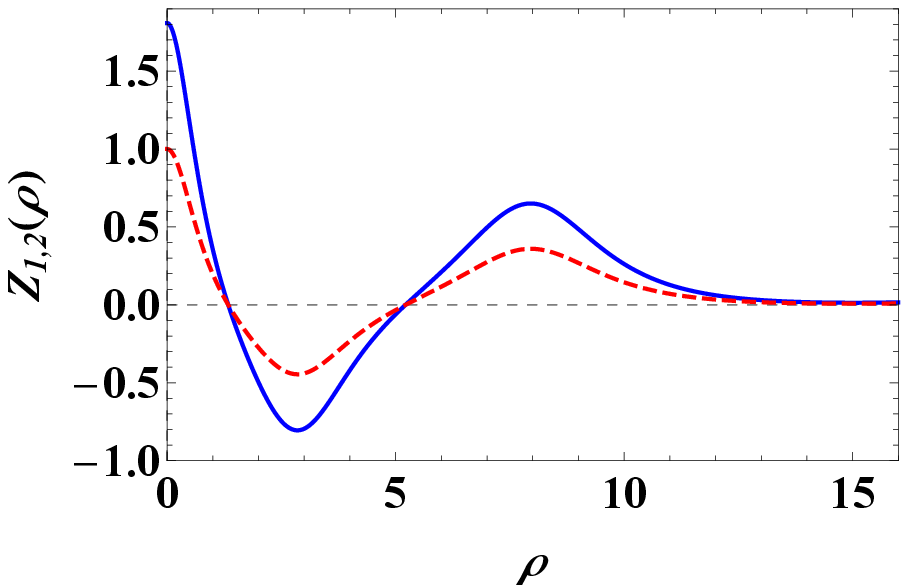}
	\caption{ (Color online) Stationary power series solutions of (\ref{ode}) with different number of nodes. Dashed (red) is $Z_1(\rho)$ and solid (blue) is $Z_2(\rho)$. The left subfigure is single-node solution with $a_0=1.3293391$, root at $r=1.85$ and the right subfigure is double-node solution with $a_0=1.8079999$, roots at $r=1.58, 5.45$.  Parameters used are: $b_0=1$, $a_1=b_1=0$, $\alpha_1=\alpha_2=0.5$, $\gamma=2$,  $n_{max}=2$, $I=5000$, and $\Delta=0.0032$.}\label{ips}
\end{figure}
Employing the algorithm  above, we obtain a family of infinite number of convergent solutions by  tuning the parameter $a_0$ and fixing the other parameters. In Fig.~\ref{ips}, we present two plots showing the single-node and double-node solutions obtained with $I=5000$ and $n_{max}=2$.
\\Although this method provides an infinite  number of non-divergent densities, due to the scalar symmetry between $\psi_1(\rho, t)$  and $\psi_2(\rho, t)$, the spin textures corresponding to these solutions which is proportional to $\psi_2/\psi_1$, turn out to be trivial.
\subsection{Numerical solutions}\label{num}

\begin{figure}[h!]
	\centering
	\includegraphics[width=0.9\linewidth]{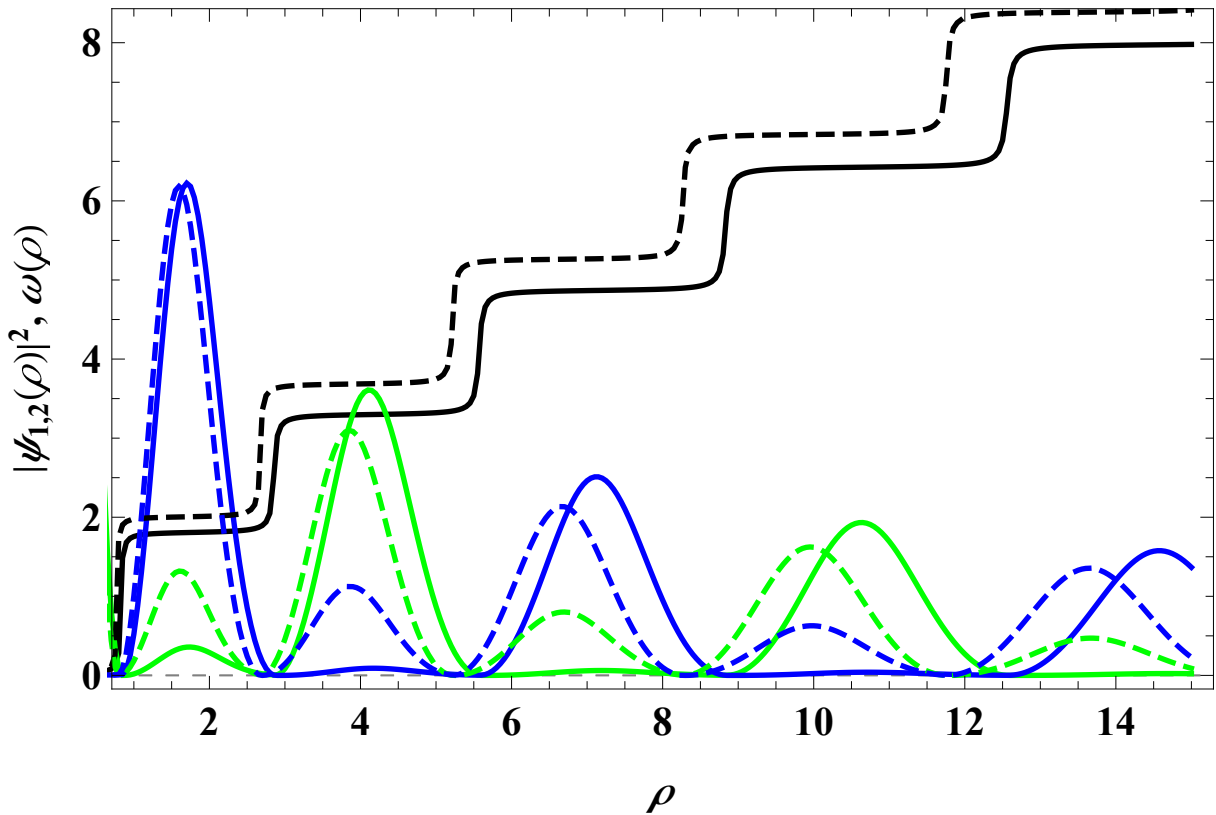}
	\caption{(Color online) The numerical solutions of the spinor
		components given by Eq.~(\ref{ansatz}) of the system (\ref{eq1}) and
		(\ref{eq2}). The solutions correspond to alternating spin-up (green) and spin-down (blue) components of the spinor wave function. The black curve corresponds to $\omega(\rho)$. The solid and dashed
		curves correspond to the circle and square in
		Fig~\ref{fig:energy}, which represent a stable and a metastable
		skyrmion, respectively. Parameters used are: $\gamma=c_1=1/2$, and
		$a_0=2.9$.}
    \label{fig:fig-psi1n-2}
\end{figure}

\begin{figure}[h]
	\centering
		\includegraphics[scale=.72]{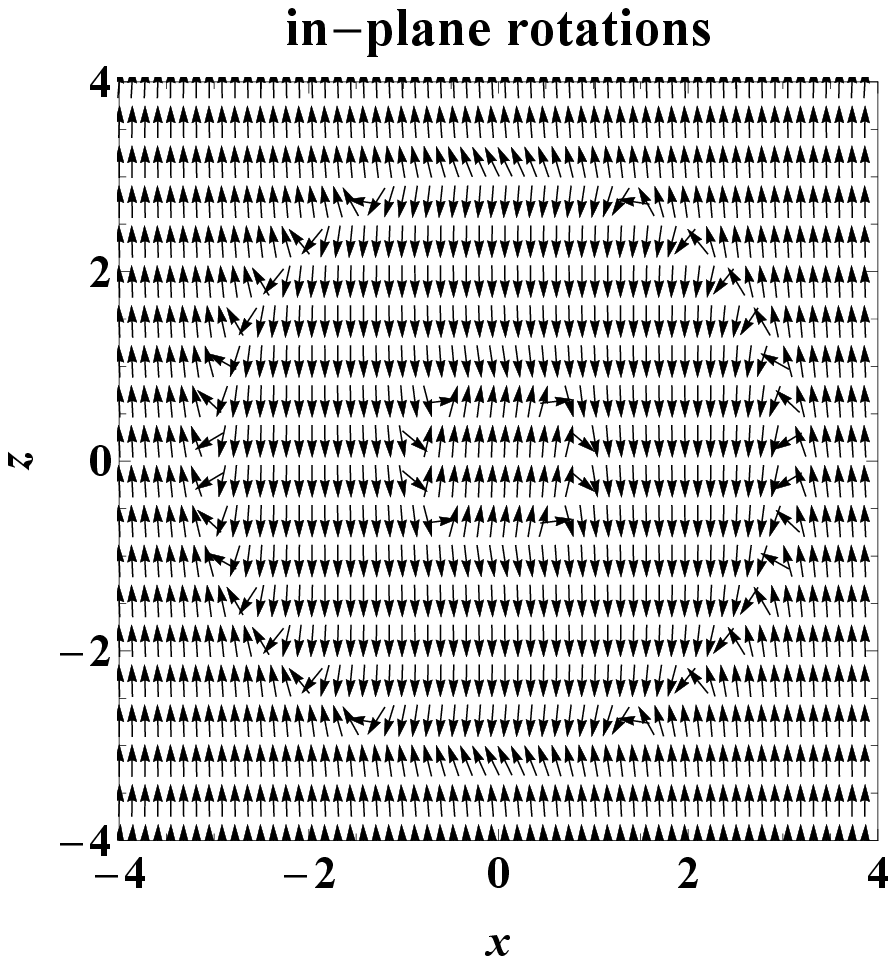}
		\includegraphics[scale=.72]{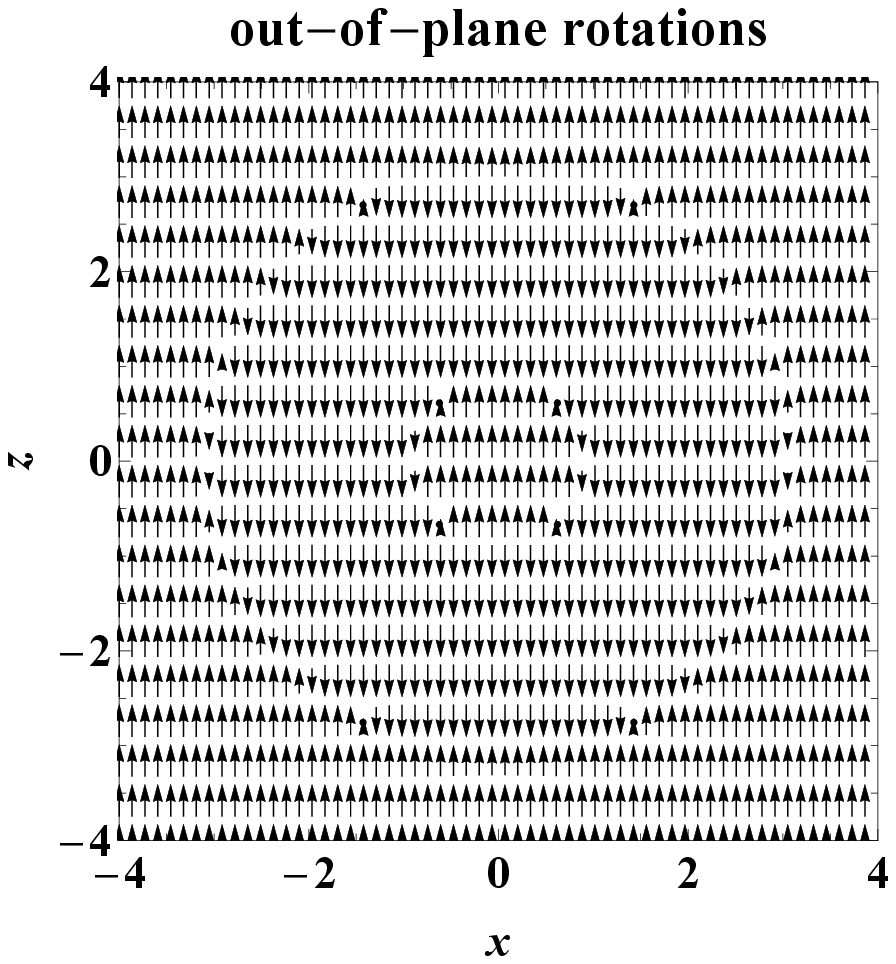}
	\caption{In-plane and out-of-plane vector representations of skyrmions in spin-1/2 system for the case of rotation in $xz$-plane around $y$-axis and $x$-axis, respectively with initial spin along $z$-axis. Parameters used are the same as in Fig. \ref{fig:fig-psi1n-2}. }
	\label{fig:3D}
\end{figure}

\begin{figure}[h]
	\centering
		\includegraphics[scale=.65]{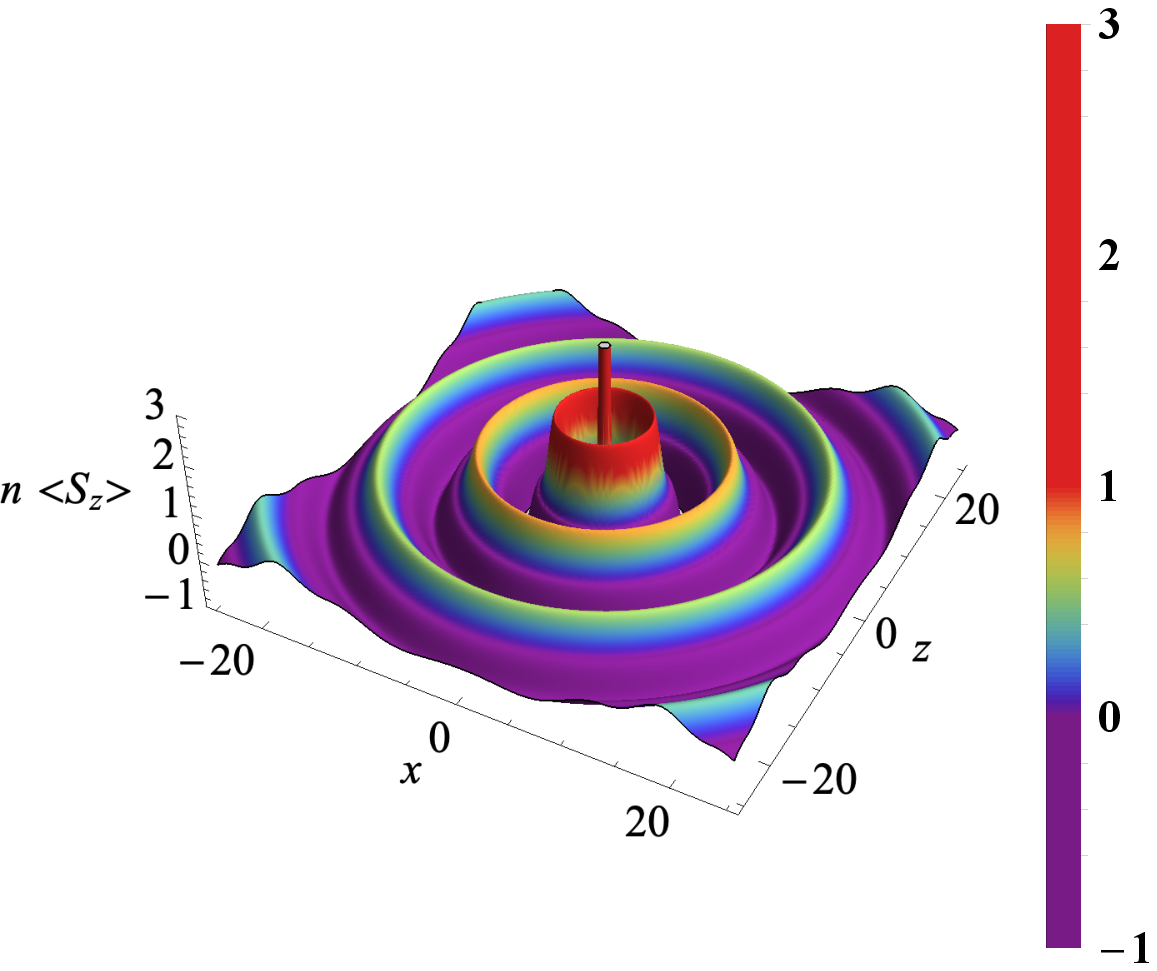}
		\includegraphics[scale=.65]{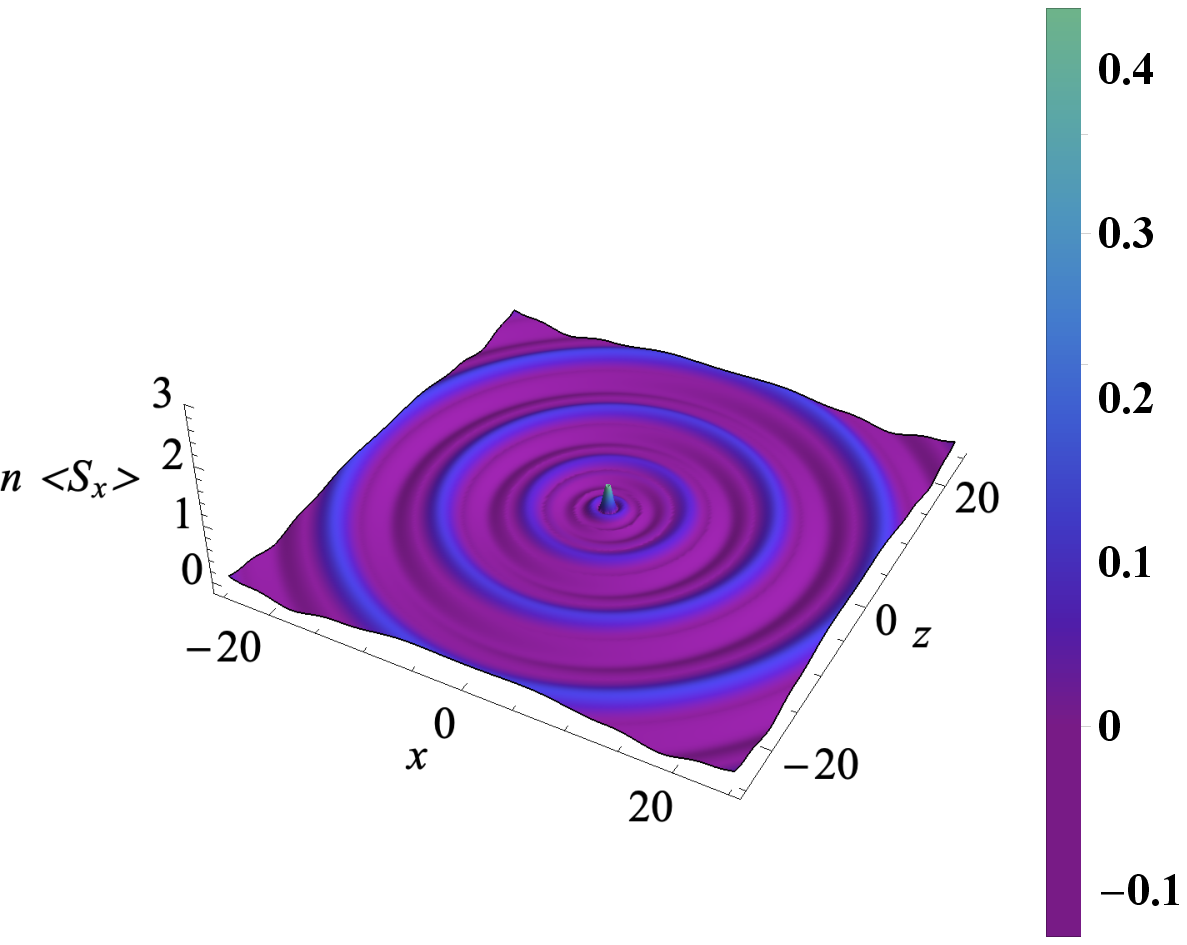}
	\caption{False-colour figures representing the average spin texture modulated by the total density, namely $n\,<S_z>$  and $n\,<S_x>$ for a skyrmion in a spin-1/2 system in $xz$-plane with initial spin along $z$-axis which is given in Table~\ref{tab:xz-plane}. The axis of rotation is $y$-axis. Parameters used are the same as in Fig. \ref{fig:fig-psi1n-2}.}
	\label{fig:vectorplot}
\end{figure}

Here, we introduce a new procedure that leads to nondiverging and
nontrivial spin textures. We start with a trial function which is
constructed on the basis of the spinor wave function for rotation
cases listed in Tables~\ref{tab:xz-plane} and \ref{tab:xy-plane}.
For instance, we consider a case of rotation from Table
\ref{tab:xz-plane} in $xz$-plane  with initial spin along $z$-axis
and $y$-axis is the axis of rotation. Our ansatz, for this case
becomes:
\begin{eqnarray}\label{ansatz}
\psi_1(\rho, t) = a(\rho) \cos [\omega (\rho)],\nonumber \\
\psi_2(\rho, t) = a(\rho) \sin [\omega (\rho)],
\end{eqnarray}
where $a(\rho) = \sqrt{n(\rho)}$. Substituting this trial function into the system given in Eq.~\eqref{MS} and then requesting the coefficients of $\cos [\omega (\rho)]$ and $\sin[\omega (\rho)]$ to vanish separately, we get two coupled equations in terms of $a(\rho)$ and $\omega(\rho)$
\begin{equation}\label{eq1}
2 a'(\rho) \omega'(\rho) + a(\rho)\Big(\dfrac{\omega'(\rho)}{\rho} + \omega''(\rho)\Big)=0,
\end{equation}
\begin{equation}\label{eq2}
\gamma~ a^3(\rho) + \dfrac{a'(\rho)}{\rho} - a(\rho)\Big( \dfrac{1}{4\rho^2} + \omega'^2(\rho) \Big) + ra'(\rho) =0.
\end{equation}
We solve Eq.~\eqref{eq1} for $\omega(\rho)$ as
\begin{equation}\label{omega}
\omega (\rho) = \int \dfrac{c_1}{\rho~ a(\rho)^2} d\rho + c_2,
\end{equation}
where $c_1$ and $c_2$ are constants of integration. By substituting the above relation for $\omega(\rho)$ in Eq.~\eqref{eq2}, our problem \eqref{MS} is reduced into the following single equation
\begin{equation}\label{reduced}
c_1^2 - \gamma ~ \rho^2~ a^6 - \rho~ a^3 \Big(a' + \rho~ a'' \Big) = 0.
\end{equation}
We solve this equation for $a(\rho)$ numerically. The initial conditions used are  $a(0) = a_0$ and $a'(0) = 0$. We choose $a_0$ as the tuning parameter for the calculation. The results are shown in Fig.~\ref{fig:fig-psi1n-2}.
Similarly, any of the rotation cases given in
Tables~\ref{tab:xz-plane} and \ref{tab:xy-plane}  can be considered
for the substitution of $\psi_1(\rho, t)$ and $\psi_2(\rho, t)$. It
turns out, however, that all cases of rotations lead to the same
Eq.~\eqref{reduced} with the same relation of $\omega(\rho)$ as
given in Eq.~\eqref{omega}.
\\
All possible cases of rotations discussed in Table
\ref{tab:xz-plane}  correspond to the two fundamental types of
skyrmions which represent rotation either in-plane or out-of-plane.
The rotation of spin around its own axis is a trivial case. The
in-plane and out-of-plane spin textures $<S_x>$ and $<S_z>$ given by the expressions
in Table~\ref{tab:xz-plane} for the cases of rotation around $y$-axis and $x$-axis, respectively,
with initial spin along $z$-axis are shown in Fig.~\ref{fig:3D}.
These results are obtained from solving the Eq.~\eqref{reduced} numerically.
The structure of $<S_y>$ is trivial (constant/plain texture) for this case. These spin textures are, however, modulated by the total density of both spin components. This kind of modulation is applicable for the adjustment of carrier distributions for current density change and light intensity \cite{24} and also for the nonlinear resonator \cite{25}.
 To show such modulations, we plot in Fig.~\ref{fig:vectorplot} the quantities $n\,<S_x>$ and $n\,<S_z>$ for the case of in-plane rotations.
In order to find the spin texture of skyrmions for the cases of
rotations around $\rho$ in the $xy$-plane, as listed in Table
\ref{tab:xy-plane}, we follow the same procedure as discussed above. However, for the case of rotation around $\rho$, the spin texture
will be dependent not only on the rotation angle $\omega(\rho)$ but also
on the projection angle $\phi$, as a result we expect fundamentally
different skyrmions. We consider the system \eqref{2DMS} to be
solved for this case which includes also $\phi$ dependence.
\begin{figure}[h!]
	\centering
	\includegraphics[width=0.9\linewidth]{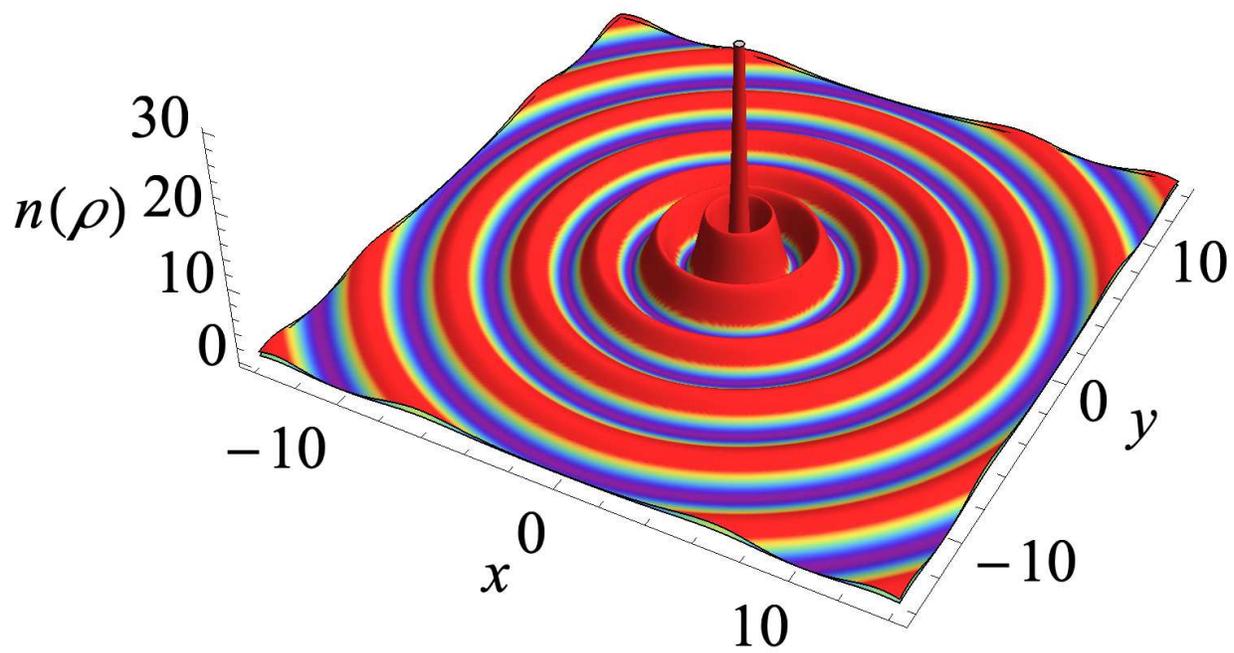}
	\caption{The total density for the case of rotation around $\rho$ in $xy$-plane with initial spin along $z$-axis.}
	\label{density}
\end{figure}
Now, we
take an example from Table \ref{tab:xy-plane} of the rotation around
$\rho$ in $xy$-plane with initial spin oriented along $z$-axis and
the trial function becomes
\begin{eqnarray}\label{ansatz2}
\psi_1(\rho, \phi, t) &=& a(\rho) ~e^{ik_1 \phi} \cos [\omega (\rho)],\nonumber \\
\psi_2(\rho, \phi, t) &=& a(\rho)~ e^{ik_2 \phi} \sin [\omega (\rho)] \times \nonumber\\ &&e^{-i(\pi/2-\phi)},
\end{eqnarray}
where $k_1$ and $k_2$ are related as $k_1=1+k_2$. The total spinor densities for this case are shown in Fig.~\ref{density}.
It is also noteworthy that all cases of rotation around $\rho$ lead to the same equation, \eqref{reduced}.  The spin texture for the case of rotation in $xy$-plane around $\rho$ with initial spin along $z$-axis is given in Fig.~\ref{fig:3dplot-last-case}. It can be seen from the figure, that the  $<S_z>$ component has the same structure as in the previous case shown in Fig.~\ref{fig:3D} because there is no $\phi$ dependence in this component.
The vector representation of the skyrmions in spin-1/2 system for the above mentioned rotation is shown below in Fig.~\ref{fig:vectorplot-last-case}.
We found two other unique skyrmion textures for the case of rotation around $\rho$ which are shown in Fig.~\ref{fig:twotextures}. The orientation of initial spin is along $y$-axis and the expressions for average spin components are given in Table~\ref{tab:xy-plane}.\\
It is clear from the expressions of average spin components given in
Table~\ref{tab:xy-plane} that there are four distinguished textures
for the case of rotation around $\rho$, as shown in
Figs.~\ref{fig:3dplot-last-case} and \ref{fig:twotextures}. For the
case of axial symmetry as discussed in Table~\ref{tab:xz-plane}, we
have only two fundamental skyrmion textures which are plotted in
Fig.~\ref{fig:3D}.
\begin{figure}[!h]
	\centering
		\includegraphics[scale=.75]{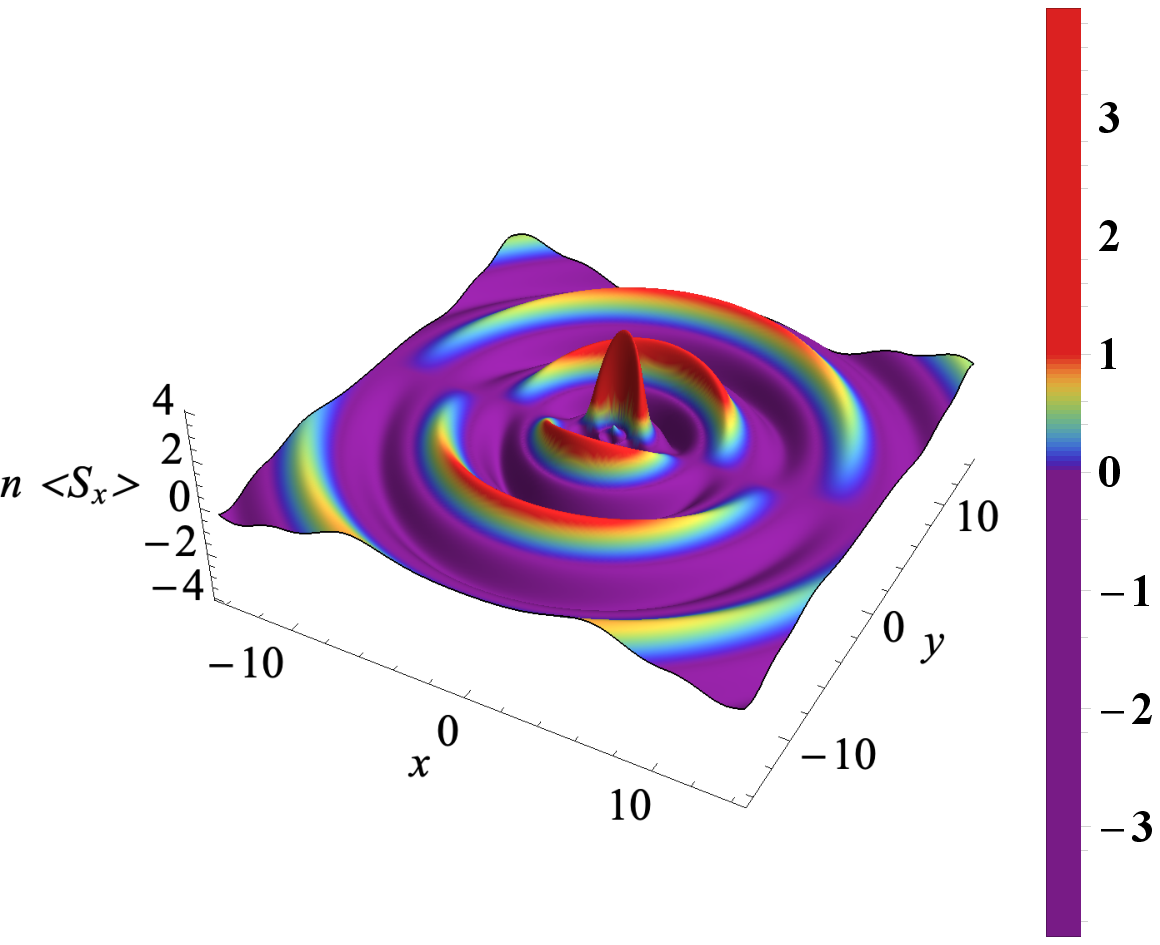}
		\includegraphics[scale=.75]{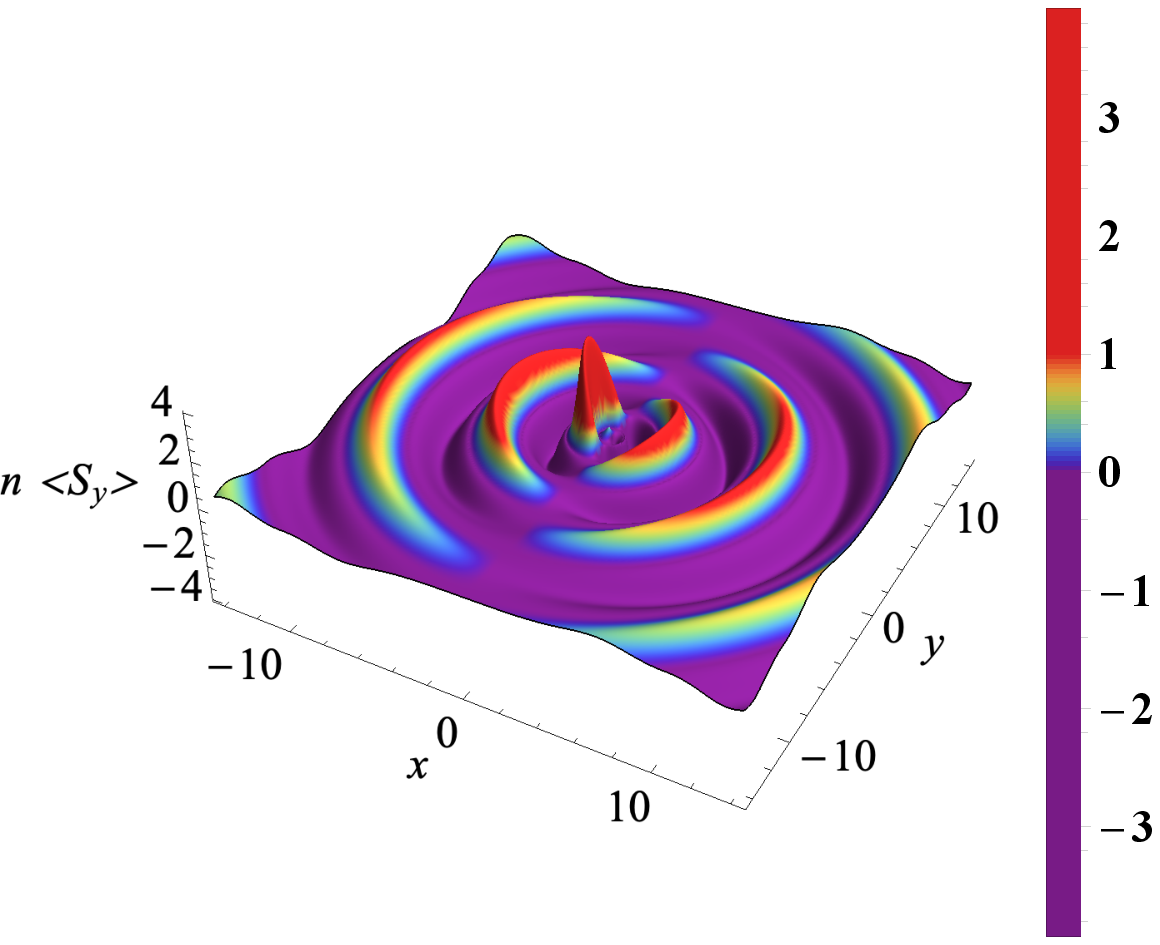}
		\includegraphics[scale=.75]{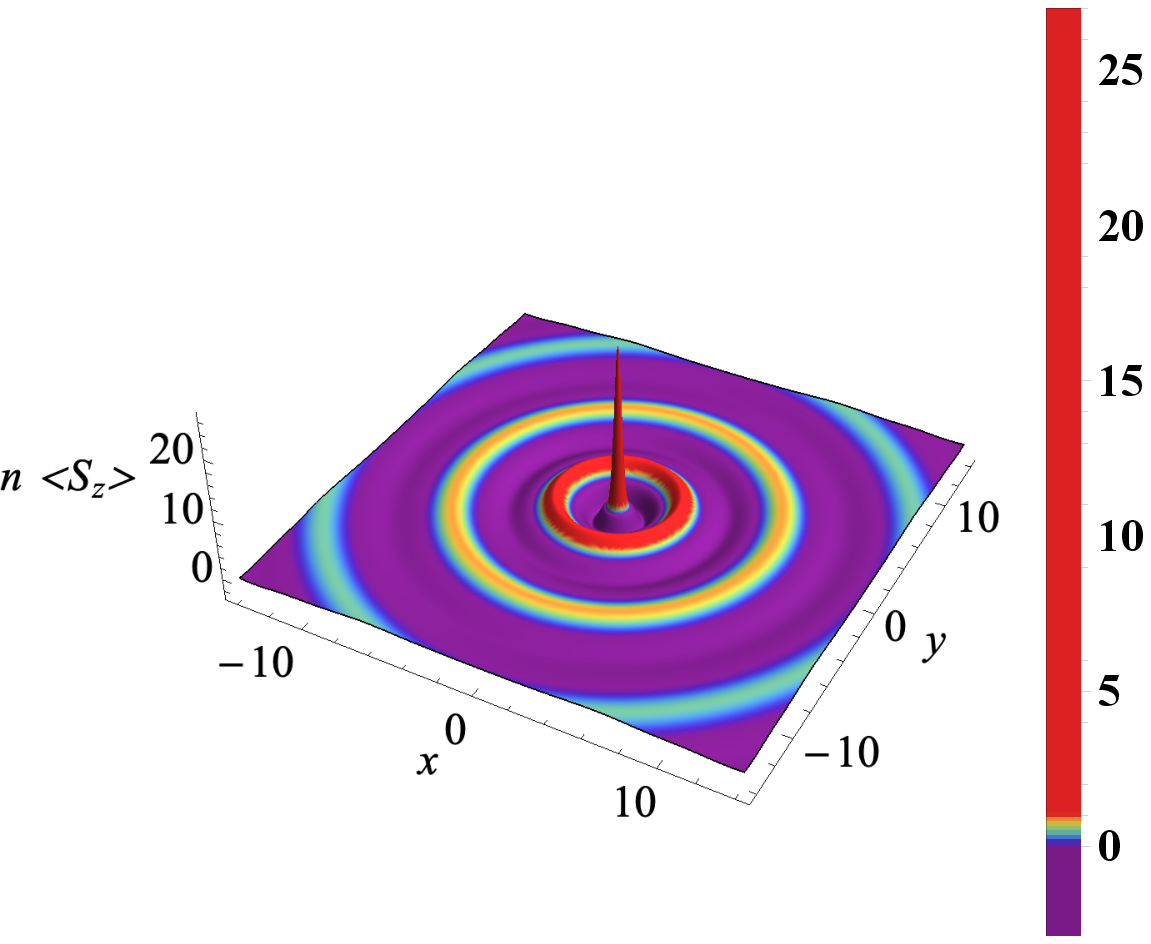}
	\caption{False-colour figures representing the average spin $n\,<S_x>$, $n\,<S_y>$, and $n\,<S_z>$ for a skyrmion in a spin-1/2 system with the rotation around $\rho$ in $xy$-plane. The initial spin orientation is along $z$-axis.}
	\label{fig:3dplot-last-case}
\end{figure}
\begin{figure}[!h]
	\centering
		\includegraphics[scale=.75]{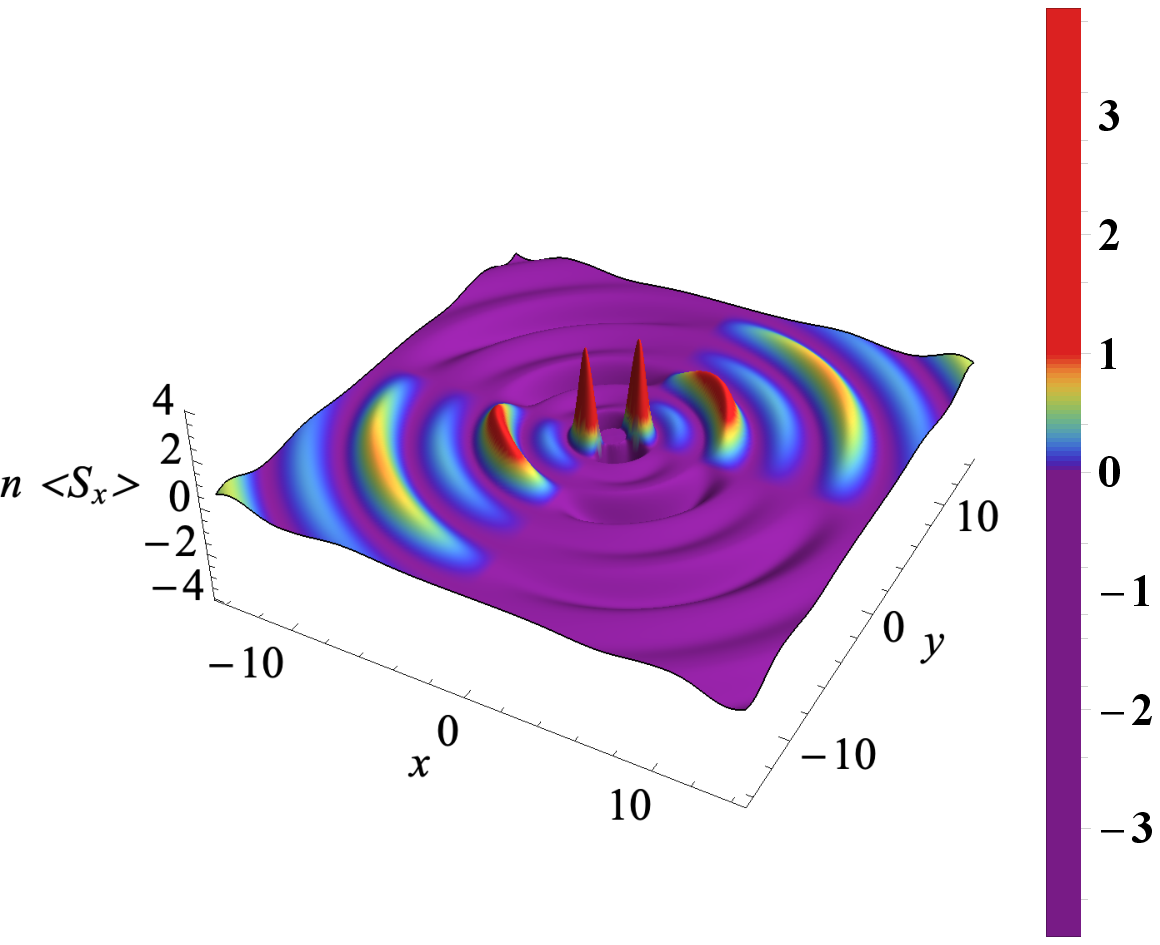}
		\includegraphics[scale=.75]{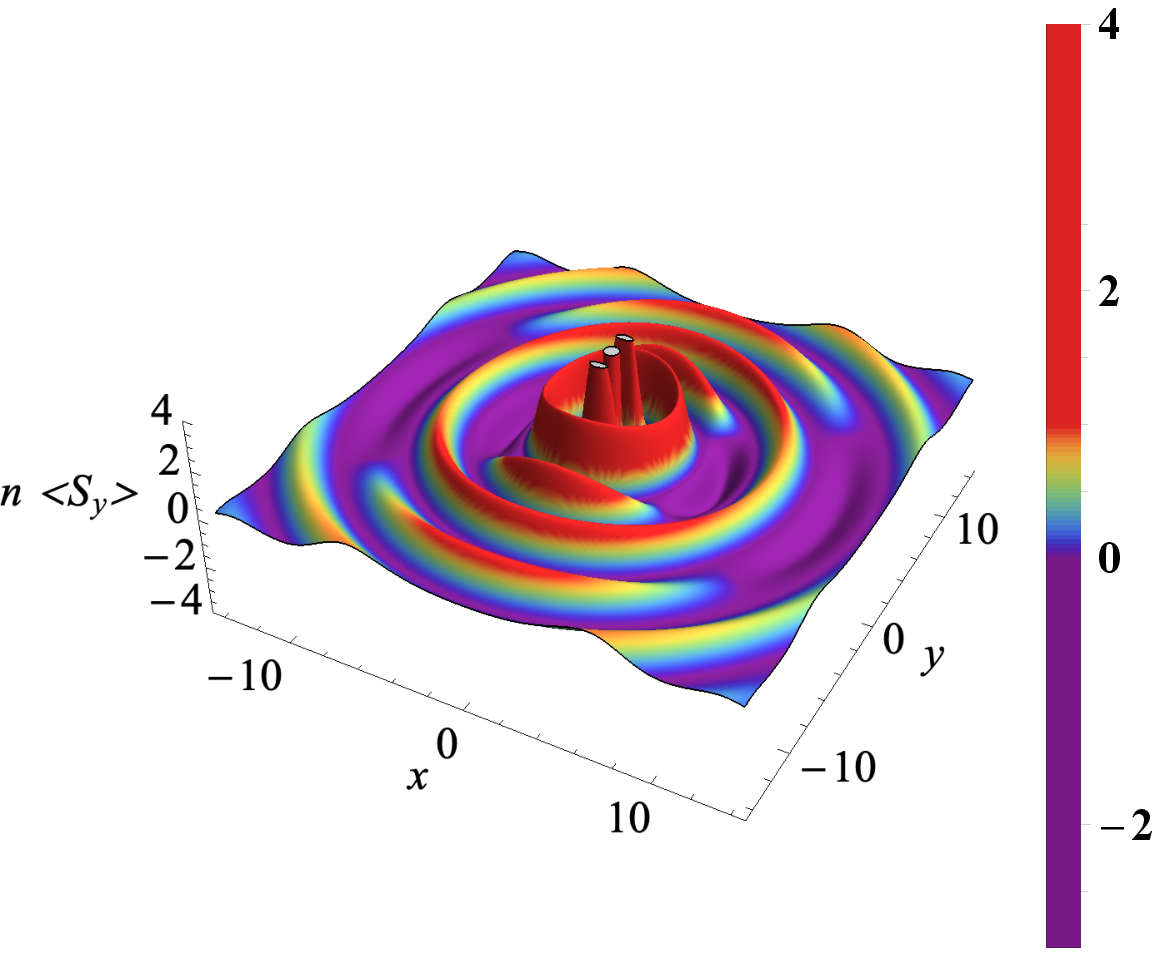}
		\includegraphics[scale=.75]{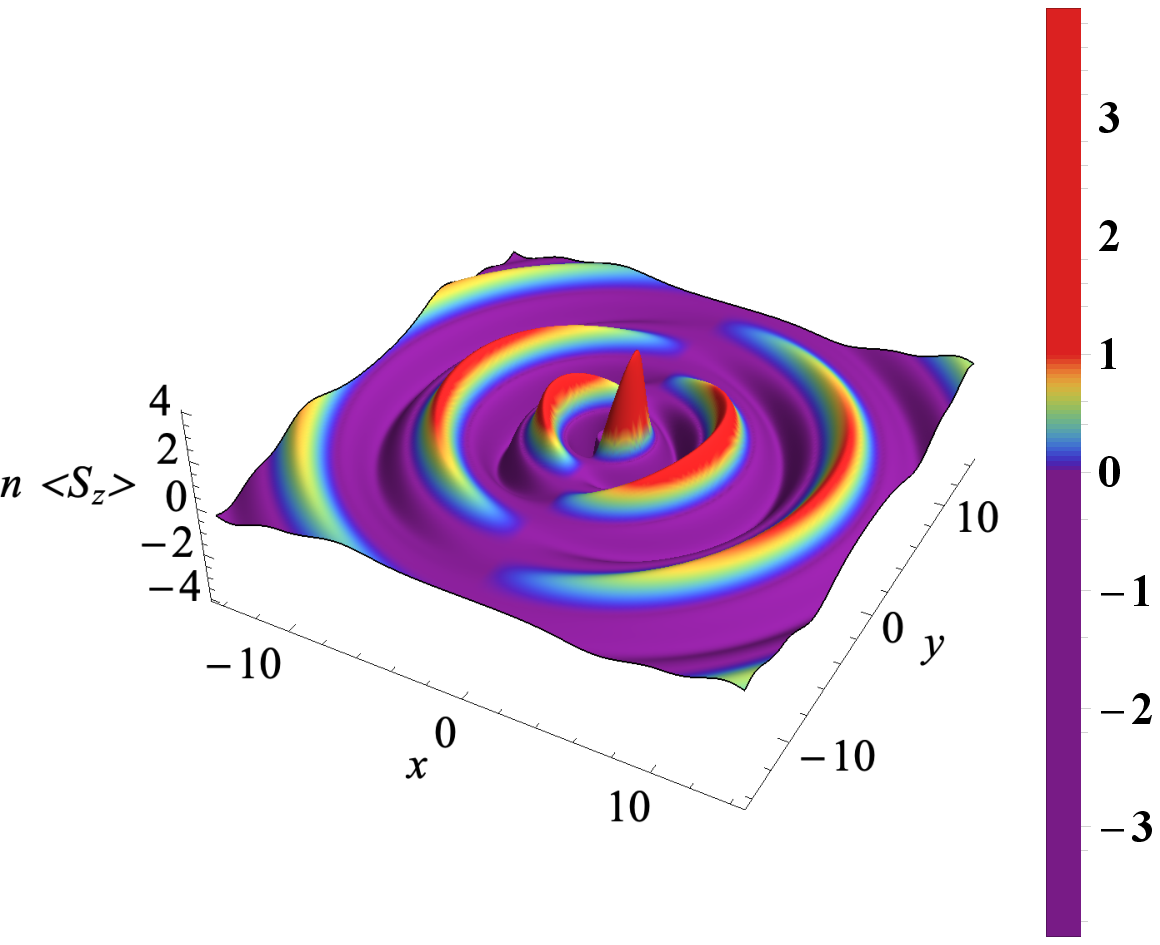}
	\caption{False-colour figures representing the average spin $n\,<S_x>$, $n\,<S_y>$, and $n\,<S_z>$ for a skyrmion in a spin-1/2 system with the rotation around $\rho$ in $xy$-plane. The initial spin orientation is along $y$-axis.}
	\label{fig:twotextures}
\end{figure}
\begin{figure}[!h]
	\centering
	\includegraphics[width=.65\linewidth]{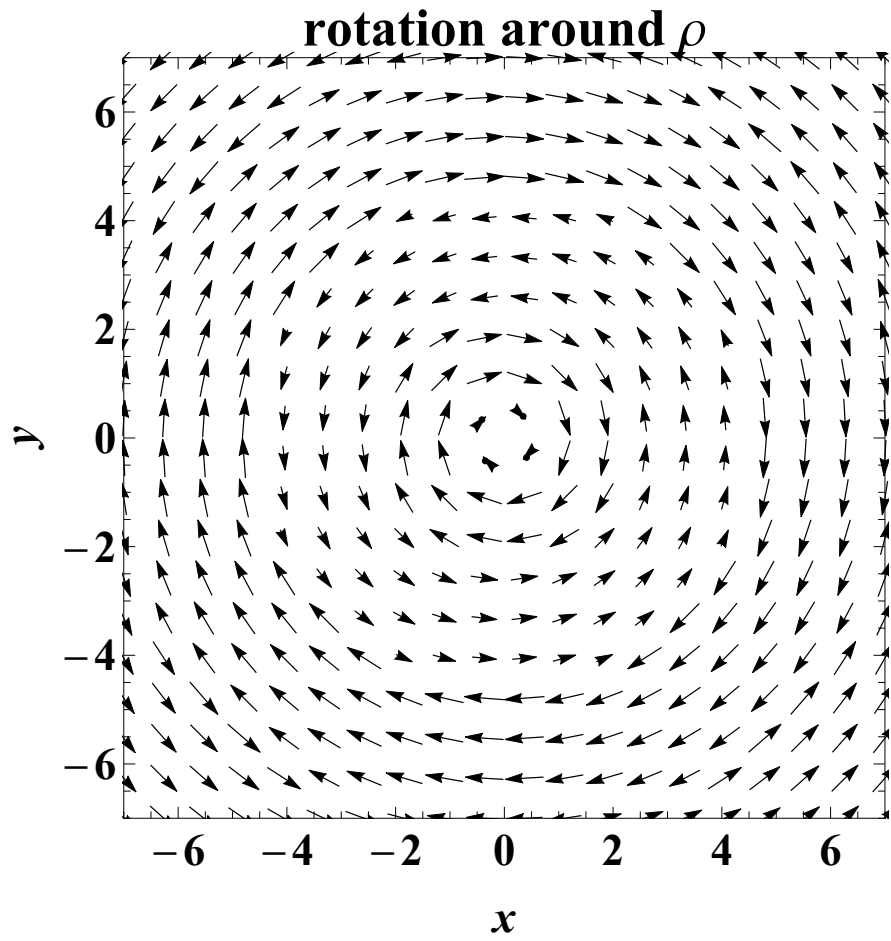}
	\vspace{.1cm}
	\caption{The vector representation of skyrmions in spin-1/2 system for the case of rotation around $\rho$ in $xy$-plane with initial spin along $z$-axis.}
	\label{fig:vectorplot-last-case}
\end{figure}

\section{Stability of the non-trivial Skyrmions}\label{stab}
In order to investigate the stability of the skyrmions, we calculated the energy functional for both cases i.e rotation around fixed axes (axial symmetry) as given in Table \ref{tab:xz-plane}  and rotations around $\rho$ which is summarized in Table \ref{tab:xy-plane}.
The energy functional corresponding to system  \eqref{2DMS} reads
\begin{eqnarray}\label{energy}
E &=& \int_{0}^{2\pi} d\phi \int_{0}^{\infty} \Big[ \alpha \Big(|\Psi_1|^2 + |\Psi_2|^2\Big)  \nonumber\\ &-&\dfrac{\gamma}{2} \Big( |\Psi_1|^2 + |\Psi_2|^2 + 2 |\Psi_1|^2 |\Psi_2|^2 \Big) + |\Psi_{1\rho}|^2    \nonumber\\ &+& |\Psi_{2\rho}|^2 +\dfrac{1}{\rho^2}\Big( |\Psi_{1\phi}|^2 + |\Psi_{2\phi}|^2\Big) \Big] \rho~ d\rho.
\end{eqnarray}
We substitute the specific trial function in Eq.~\eqref{energy} to
find the expression for the energy functional of that specific case
of rotation. For instance, we substitute the trial function
\eqref{ansatz2} for the case of rotation around $\rho$ in $xy$-plane
with initial spin along $z$-axis in above relation and find the
following expression
\begin{eqnarray}
E &=& \int \Big[-\dfrac{1}{2} \gamma~ a(\rho)^4 + a'(\rho)^2 + \dfrac{c_1^4}{\rho^2~a(\rho)^2} + \nonumber\\ &&   \alpha ~a(\rho)^2 + \dfrac{a(\rho)^2 (1+2k_2 +k_2^2)}{\rho^2} \Big]\rho~d\rho,
\end{eqnarray}
where $k_1 = 1 + k_2$.\\
While considering axial symmetry \eqref{MS}, for example,
rotation around $y$-axis with initial spin along $z$-axis,
we use the trial function \eqref{ansatz} into \eqref{energy}
and obtain the following result for energy functional
\begin{eqnarray}
E &=&\int \Big[-\dfrac{1}{2}~ \gamma~ a(\rho)^4 + a'(\rho)^2 + \alpha ~a(\rho)^2 \nonumber\\ &+& \dfrac{c_1^4}{\rho^2~a(\rho)^2}\Big] \rho~ d\rho .
\end{eqnarray}
We find a global minimum in the energy functional as well as many
local minima. The local minima correspond to a state of concentric
rings with spins alternating sharply between 1/2 and -1/2. The total
density within a ring is contributed by only one component of spin,
either spin up or spin down. On the other hand, mixed states of spin
in which the total density is contributed by both component of spin,
i.e spin up and spin down correspond to a metastable skyrmion. The
profile of the two spin componenets corresponding to a stable and a
metastable skyrmion (circle and square on the energy curve) are
shown in Fig.~\ref{fig:fig-psi1n-2}.
\begin{figure}[!h]
	\centering
	\includegraphics[width=0.9\linewidth]{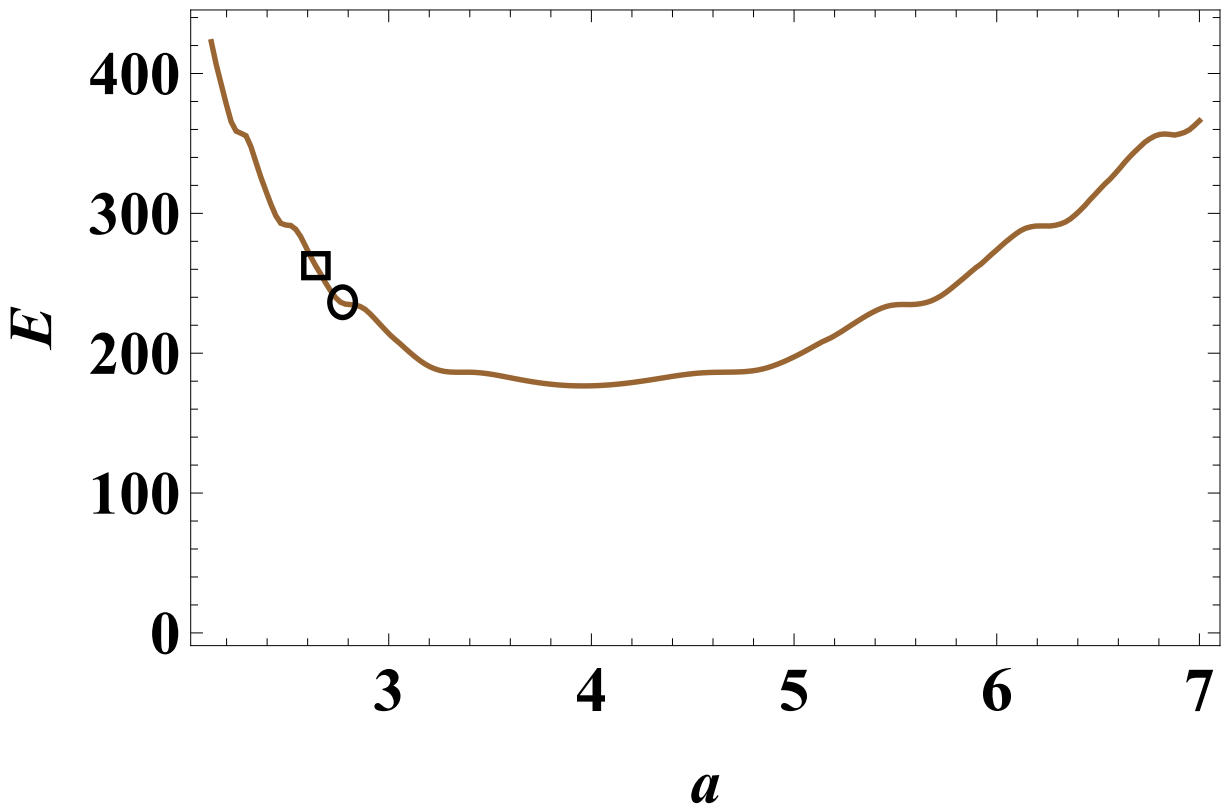}
	\caption{The energy for the case of rotation around $y$-axis with initial spin along $z$-axis.}
	\label{fig:energy}
\end{figure}

\section{Conclusion}\label{con}
We mapped the spin-1/2 system to a 2D Manakov system through
a rotation operator that gives the spin texture of skyrmions. We
have investigated all possible 2D skyrmion textures, as listed in
Tables 1 and 2.
We solved the 2D Manakov system using various analytical and numerical
methods. While the similarity transformation method maps all solutions of the integrable 1D Manakov system to the 2D Manakov system, the solutions of the latter turn out to diverge at $\rho=0$.
Nondiverging solutions were then obtained using a power series method.
However, the spin texture associated with these solutions turned out to
be trivial, i.e., no texture. Finally, we considered a numerical solution of a
system of coupled equations for the skyrmion density, $n(\rho,\phi)$, and texture,
$\omega(\rho,\phi)$. This led to nondiverging and nontrivial spin textures.
Then, we investigated the stability of these nontrivial nondiverging skyrmions
by calculating their energy functional in terms of their effective size.
It turned out that stable skyrmions correspond to concentric rings of spin
components alternating between spin up and spin down. Metastable
states, where energy is either increasing or decreasing with skyrmion size, correspond
to concentric rings of mixed spin components.
\\
\\Our results show that, in contrast with the established fact that in
two dimensions localized solutions of the NLSE are unstable, the
two spin states stabilize each other against collapse and allow for nontrivial
stable two-dimensional topological excitations. Our results are also applicable to doubly polarized optical pulses. We strongly believe that this work is an important addition to the effort of realizing topological excitations.

\appendix
\numberwithin{equation}{section}
\numberwithin{figure}{section}
\section{Solutions of the 2D Manakov system}\label{appa}
\begin{figure}[!h]
	\centering
	\includegraphics[width=1\linewidth]{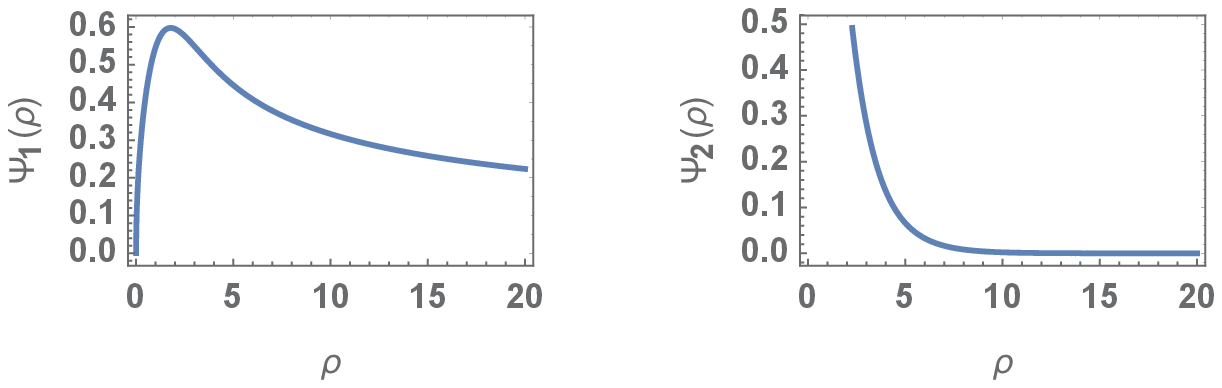}
	\caption{The graphical representation of $\psi_1(\rho)$ and $\psi_2(\rho)$ for solution-1.}
	\label{fig:sol9-plot}
	\label{fig:11}
\end{figure}

Using the similarity transformation described in Sec.~\ref{sim}, we found many new solutions for the 2D Manakov system \eqref{2DMS}, here we mentioned only two of them for their significance. The full list of solution is compiled by Ref.~\cite{26}. \\ \\
\textbf{Solution-1}
\begin{align}
\psi_1(\rho,t) = \frac{1}{\sqrt{\rho}}{\tanh \left( \sqrt{\dfrac{3}{8}}~\rho \right)} ~e^{ -i(1-3t)} \\
\psi_2(\rho,t) =\sqrt{\frac{5}{2}} \frac{1}{\sqrt{\rho}}{\sech
	\left( \sqrt{\dfrac{3}{8}}~\rho \right)} ~e^{\frac{15}{8} i(-1+t)}
\end{align}
The choice of parameters are $a_{11}=-1$, $a_{12}=2$, $a_{21}=1$, and
$a_{13}=a_{22}=a_{23}=\frac{1}{2}$. See Fig.~\ref{fig:11}.

\begin{figure}[!h]
	\centering
	\includegraphics[width=1\linewidth]{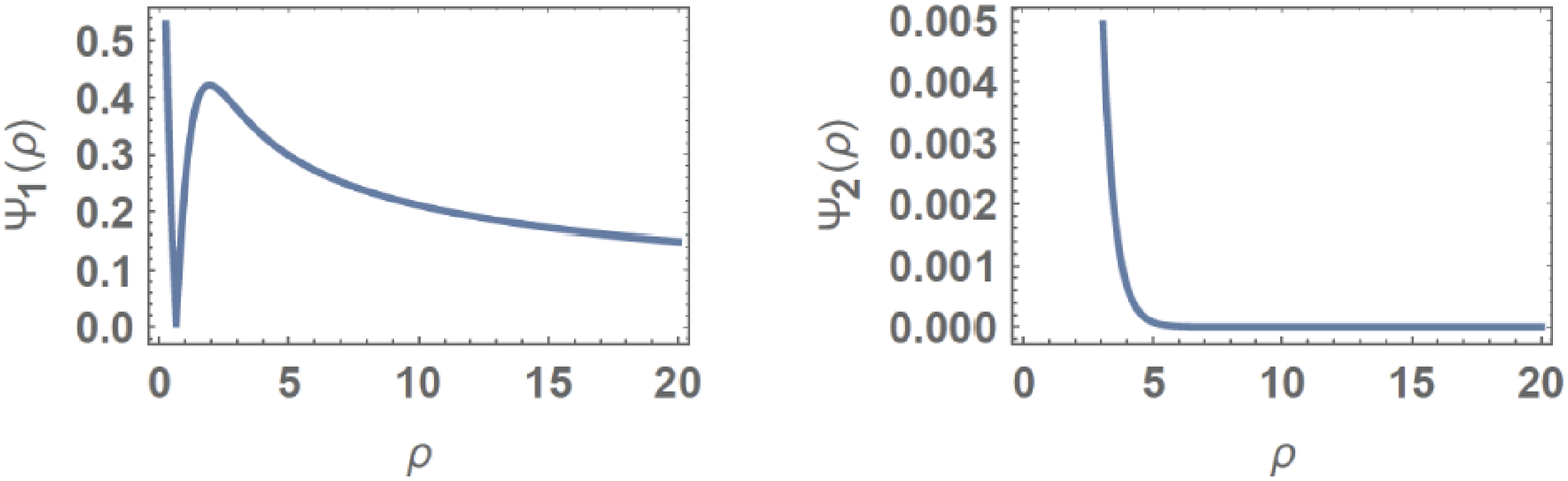}
	\caption{The graphical representation of $\psi_1(\rho)$ and $\psi_2(\rho)$ for solution-2.}
	\label{fig:12}
\end{figure}

\noindent\textbf{Solution-2}
\begin{align}
\psi_1(\rho,t) = \frac{1}{3 \sqrt{\rho}}\Big(-2+3 \sech^2 [\rho]\Big) ~e^{ 2i(1-t)} \\
\psi_2(\rho,t) = \frac{1}{\sqrt{\rho}}~{\sech^2 [\rho]} ~e^
{-2i(1-t)}
\end{align}
The parameters are $a_{11}=a_{21}=1$, $a_{12}=a_{22}=-\frac{9}{2}$, and
$a_{13}=a_{23}=\frac{9}{2}$. See Fig.~\ref{fig:12}.

\section{Similarity transformation}\label{appb}
The results of all the unknown quantities in \eqref{trans} and all the coefficients in \eqref{source} are listed below:\\ \\
\textbf{For $\bf{\Psi_1 (\vec{r},t)}$:}\\ \\
$ T(\rho,t)= g_1(t),\\
p_1(\rho,t) = \dfrac{1}{e^{i B_1(\rho,t)} ~A(\rho,t) ~g_1^{'}(t)},\\
b_{11}(\rho,t) =   \dfrac{a_{11}~ g_1^{'}(t) }{P^2_\rho (\rho,t)},\\
b_{12}(\rho,t) =   \dfrac{a_{12}~ g_1^{'}(t) }{A^2 (\rho,t)},\\
b_{13}(\rho,t) =   \dfrac{a_{13}~ g_1^{'}(t) }{A^2 (\rho,t)},\\
A(\rho,t) =   \dfrac{ g_2(t) }{\sqrt{\rho P_\rho(\rho,t)}},\\
B_1(\rho,t) = - \int{       \dfrac{P_t(\rho,t) P_\rho(\rho,t) }{2 ~a_{11} ~ g_1^{'}(t)}} d\rho+ g_3(t),\\
b_{14i}(\rho,t) = - \dfrac{g_2^{'}(t)}{g_2(t)} +\dfrac{P_{\rho t}
	(\rho,t)}{P_\rho(\rho,t)}$,
\begin{eqnarray}
b_{14r}(\rho,t) &=&  \dfrac{1}{4~ a_{11}~ g_1^{'2}(t)}\Big[2\int
P_t(\rho,t) P_\rho (r,t) d\rho~ g_1^{''}(t) \nonumber\\&-&
g_1^{'}(t)\Big(2\int(I ) d\rho + P_t^2(\rho,t) -\nonumber\\&&
\dfrac{N}{\rho^2 P_\rho ^4 (\rho,t)}\Big) \Big]+g_3^{'}(t).
\end{eqnarray}
where $I=P_{tt}(\rho,t) P_\rho(\rho,t) + P_t(\rho,t)
P_{\rho t}(\rho,t)$, and
$N=a_1^2~ g_1^{'2}(t)[ P_\rho^2(\rho,t)+3 \rho^2 ~P_{\rho
	\rho}^2(\rho,t) - 2 \rho^2 P_\rho(\rho,t) P_{\rho \rho
	\rho}(\rho,t)]$.\\ \\
\textbf{For $\bf{\Psi_2 (\vec{r},t)}$:}\\ \\
$ T(\rho,t)= g_1(t),\\
p_2(\rho,t) = \dfrac{1}{e^{i B_2(\rho,t)} ~A(\rho,t) ~g_1^{'}(t)},\\
b_{21}(\rho,t) =   \dfrac{a_{21}~ g_1^{'}(t) }{P^2_\rho (\rho,t)},\\
b_{22}(\rho,t) =   \dfrac{a_{22}~ g_1^{'}(t) }{A^2 (\rho,t)},\\
b_{23}(\rho,t) =   \dfrac{a_{23}~ g_1^{'}(t) }{A^2 (\rho,t)},\\
A(\rho,t) =   \dfrac{ g_2(t) }{\sqrt{\rho P_\rho(\rho,t)}},\\
B_2(\rho,t) = - \int{       \dfrac{P_t(\rho,t) P_\rho(\rho,t) }{2 ~a_{21} ~ g_1^{'}(t)}} d\rho+ g_3(t),\\
b_{24i}(\rho,t) = - \dfrac{g_2^{'}(t)}{g_2(t)} +\dfrac{P_{\rho t}
	(\rho,t)}{P_\rho(\rho,t)}$,
\begin{eqnarray}
b_{24r}(\rho,t) &=&  \dfrac{1}{4~ a_{21}~ g_1^{'2}(t)}\Big[2\int
P_t(\rho,t) P_\rho (r,t) d\rho~ g_1^{''}(t) \nonumber\\&-&
g_1^{'}(t) \Big(2\int(I) d\rho  + P_t^2(\rho,t) - \nonumber\\&&\dfrac{ N}{\rho^2 P_\rho ^4
	(\rho,t)}\Big) \Big]+g_3^{'}(t).
\end{eqnarray}
where $I = P_{tt}(\rho,t) P_\rho(\rho,t) + P_t(\rho,t) P_{\rho t}(\rho,t)$, and $N = a_1^2~ g_1^{'2}(t)[P_\rho^2(\rho,t)+3 \rho^2 ~P_{\rho \rho}^2(\rho,t) - 2 \rho^2
P_\rho(\rho,t) P_{\rho \rho \rho}(\rho,t)]$.
Here $a_{11}$, $a_{12}$, $a_{13}$, $a_{21}$, $a_{22}$ and $a_{23}$ are all arbitrary real constants.

\section*{Acknowledgment} The authors acknowledge the support of UAE University through grants UAEU-UPAR(4) 2016 and UAEU-UPAR(6) 2017.


\begin{thebibliography}{1}
	
	\bibitem{1}
	 S. V. Manakov, ``On the theory of two-dimensional stationary self-focusing of electromagnetic waves,"  \textit{J. Exp. Theor. Phys.}, vol. 38, no. 2, pp. 248-253, 1974.
	
	 \bibitem{2}
	  M. R. Gupta, B. K. Som, and B. Dasgupta, ``Coupled nonlinear Schr{\"o}dinger equations for Langmuir and elecromagnetic waves and extension of their modulational instability domain,'' \textit{J. Plasma Phys.}, vol. 25, no. 3, pp. 499-507, 1981.
	
	  \bibitem{3}
	   Y. Chen and H. A. Haus, ``Solitons and polarization mode dispersion,'' \textit{Opt. Lett.}, vol. 25, no. 5, pp. 290-292, 2000.\\
	   F. K. Abdullaev, B. A. Umarov, M. R. B. Wahiddin, and D. V. Navotny, ``Dispersion-managed solitons in a periodically and randomly inhomogeneous birefringent optical fiber,'' \textit{J. Opt. Soc. Am.}, vol. 17, no. 7, pp. 1117-1124, 2000.

     \bibitem{22}
     C. Xie, M. Karlsson, P. A. Andrekson, and H. Sunnerud, ``Statistical analysis of soliton robustness to polarisation-mode dispersion,'' \textit{Electron. Lett.}, vol. 36, no. 18, pp. 1575-1577, 2000.

     \bibitem{4}
     J. U. Kang, G. I. Stegeman, J. S. Aitchison, and N. Akhmediev, ``Observation of Manakov spatial solitons in AlGaAs planar waveguides,'' \textit{Phys. Rev. Lett.}, vol. 76, no. 20, p. 3699, 1996.
	
	\bibitem{23}
	K. Xu, Y. Chen, T. A. Okhai, and L. W. Snyman, ``Micro optical sensors based on avalanching silicon light-emitting devices monolithically integrated on chips,'' \textit{Opt. Mater. Express.}, vol. 9, no. 10, pp. 3985-97, 2019.
	
		\bibitem{5}
		M. Tajiri and M. Hagiwara, ``Similarity solutions of the two-dimensional coupled nonlinear Schr{\"o}dinger equation,'' \textit{J. Phys. Soc. Jpn.}, vol. 52, no. 11, pp. 3727-3734, 1983.
		
		\bibitem{6}
		E. Ar\'{e}valo, ``Solitary wave solutions as a signature of the instability in the discrete nonlinear Schr{\"o}dinger equation,'' \textit{Phys. Rev. Lett.}, vol. 102, no. 22, p. 224102, 2009.

	\bibitem{7}
	H.Q. Zhang, X.H. Meng, T. Xu, L. L. Li, and B. Tian, ``Interactions of bright solitons for the (2+ 1)-dimensional coupled nonlinear Schr{\"o}dinger equations from optical fibres with symbolic computation,'' \textit{Phys. Scr.}, vol. 75, no. 4, p. 537, 2007.
	
	\bibitem{8}
	Y. P. Wang, B. Tian, W. R. Sun, and D. Y. Liu, ``Analytic study on the mixed-type solitons for a (2+ 1)-dimensional N-coupled nonlinear Schr{\"o}dinger system in nonlinear optical-fiber communication,'' \textit{Commun. Nonlinear Sci. Numer. Simul.}, vol. 22, no. 1-3, pp. 1305-1312, 2015.
	
	\bibitem{9}
	Y. J. Cai, C. L. Bai, and Q. L. Luo, ``Exact soliton solutions for the (2+ 1)-dimensional coupled higher-order Nonlinear Schr{\"o}dinger equations in birefringent optical-fiber communication,'' \textit{Commun. Theor. Phys.}, vol. 67, no. 3, p. 273, 2017.
	
	\bibitem{10}
	S. T. Ji and X. S. Liu, ``Generating ring dark solitons in two-component Bose--Einstein condensates,'' \textit{Phys. Lett. A}, vol. 378, no. 5-6, pp. 524-528, 2014.
	
	\bibitem{11}
	J. Stockhofe, P. G. Kevrekidis, D. J. Frantzeskakis, and P. Schmelcher, ``Dark--bright ring solitons in Bose--Einstein condensates,'' \textit{J. Phys. B}, vol. 44, no. 19, p. 191003, 2011.
	
		\bibitem{12}
		J. Hudock, P. G. Kevrekidis, B. A. Malomed, and D. N. Christodoulides, ``Discrete vector solitons in two-dimensional nonlinear waveguide arrays: Solutions, stability, and dynamics,'' \textit{Phys. Rev. E }, vol. 67, no. 5, p. 056618, 2003.
	
		\bibitem{13}
		J. W. Fleischer, M. Segev, N. K. Efremidis, and D. N. Christodoulides, ``Observation of two-dimensional discrete solitons in optically induced nonlinear photonic lattices,'' \textit{Nature}, vol. 422, no. 6928, pp. 147-150, 2003.
	
		\bibitem{14}
		H. N. Hassan and M. A. El-Tawil, ``Solving cubic and coupled nonlinear Schr{\"o}dinger equations using the homotopy analysis method,'' \textit{J. Appl. Math. Mech.}, vol. 7, no. 8, pp. 41-64, 2011.
	
		\bibitem{15}
		F. Kh Abdullaev and E. N. Tsoy, ``The evolution of optical beams in self-focusing media,'' \textit{Physica D}, vol. 161, no. 1-2, pp. 67-78, 2002.
	
		\bibitem{16}
		U. Al Khawaja and H. T. C. Stoof, ``Skyrmion physics in Bose-Einstein ferromagnets,'' \textit{Phys. Rev. A}, vol. 64, no. 4, p. 043612, 2001.
		
		\bibitem{17}
		U. Al Khawaja and H. T. C. Stoof, ``Skyrmions in a ferromagnetic Bose--Einstein condensate,'' \textit{Nature}, vol. 411, no. 6849, p. 918, 2001.
		
		\bibitem{18}
		H. T. C. Stoof, E. Vliegen, and U. Al Khawaja, ``Monopoles in an antiferromagnetic Bose-Einstein condensate,'' \textit{Phys. Rev. Lett.}, vol. 87, no. 12, p. 120407, 2001.
	
	\bibitem{19}
	J. Noh, W. A. Benalcazar, S. Huang, M. J. Collins, K. P Chen, T. L. Hughes, and M. C. Rechtsman, ``Topological protection of photonic mid-gap defect modes,'' \textit{Nat. Photonics}, vol. 12, no. 7, pp. 408-415, 2018.
	
	\bibitem{20}
	Y. Ke, X. Qin, F. Mei, H. Zhong, Y. S. Kivshar, and C. Lee, ``Topological phase transitions and thouless pumping of light in photonic waveguide arrays,'' \textit{Laser Photonics Rev.}, vol. 10, no. 6, pp. 995-1001, 2016.
	
	\bibitem{21}
	L. Y. Al Sakkaf, Q. M. Al-Mdallal, and U. Al Khawaja, ``A Numerical Algorithm for Solving Higher-Order Nonlinear BVPs with an Application on Fluid Flow over a Shrinking Permeable Infinite Long Cylinder,'' \textit{Complexity}, vol. 2018, pp. 1-11, 2018.
	
	\bibitem{24}
	K. Xu, ``Silicon MOS optoelectronic micro-nano structure based on reverse-biased PN junction,'' \textit{Phys. Status Solidi A}, vol. 216, no. 7, p. 1800868, 2019.
	
	\bibitem{25}
	A. Leuch, L. Papariello, O. Zilberberg, C. L. Degen, R. Chitra, and A. Eichler, ``Parametric symmetry breaking in a nonlinear resonator,'' \textit{Phys. Rev. Lett.}, vol. 117, no. 21, p. 214101, 2016.
	
	\bibitem{26}
	U. Al Khawaja and L. Al Sakkaf, \textit{Handbook of Exact Solutions to the Nonlinear Schr{\"o}dinger Equations}, London, IOP publishing, 2019.
		
\end{thebibliography}
\end{document}